\documentclass{aa}

\usepackage{graphicx}
\usepackage{bm}
\usepackage[varg]{txfonts}
\usepackage{multirow,xspace}
\usepackage{mathtools}
\usepackage[colorlinks=true,linkcolor=cyan,citecolor=cyan,urlcolor=cyan]{hyperref}

\graphicspath{{./}{./figs/}}

\newcommand{\sub}[1]{_{\rm #1}}

\newcommand{\dd}{\mathrm{d}}

\newcommand{\SigdotPE}{\dot{\Sigma}_{\rm PEW}}
\newcommand{\SigdotX}{\dot{\Sigma}_{\rm X}}
\newcommand{\SigdotEUV}{\dot{\Sigma}_{\rm EUV}}
\newcommand{\SigdotDW}{\dot{\Sigma}_{\rm MDW}}
\newcommand{\rhomid}{\rho\sub{mid}}
\newcommand{\cs}{c\sub{s}}
\newcommand{\arp}{\overline{\alpha_{r\phi}}}
\newcommand{\apz}{\overline{\alpha_{\phi z}}}

\newcommand{\Mstar}{M_\star}
\newcommand{\Lstar}{L_\star}

\newcommand{\Msun}{M_{\sun}}

\newcommand{\Lsun}{L_{\sun}}

\newcommand{\amu}{m_\mathrm{u}}
\newcommand{\kB}{ k\sub{B}}
\newcommand{\au}{{\rm au}}

\newcommand{\K}{{\rm K}}
\newcommand{\Tmid}{T\sub{mid}}
\newcommand{\Tirr}{T\sub{irr}}
\newcommand{\Tacc}{T\sub{acc}}

\newcommand{\cw}{C\sub{w}}
\newcommand{\Cw}{C\sub{w}}
\newcommand{\cwo}{C\sub{w,0}}
\newcommand{\cwe}{C\sub{w,e}}
\newcommand{\erad}{\epsilon\sub{rad}}
\newcommand{\Frad}{F\sub{rad}}

\newcommand{\Grp}{\Gamma_{r\phi}}
\newcommand{\Gpz}{\Gamma_{\phi z}}

\renewcommand{\d}{\mathrm{d}}
\newcommand{\pr}[1]{\ensuremath{\left(#1\right)} }
\newcommand{\pd}[2]{\ensuremath{\dfrac{\partial #1}{\partial #2} }}
\newcommand{\pf}[2]{\ensuremath{\left( \frac{#1}{#2} \right)} }
\newcommand{\rin}{r_{\rm in}}
\newcommand{\fheat}{\ensuremath{f_{\rm heat}}}
\newcommand{\sgmheat}{\ensuremath{\Sigma_{\rm heat}}}
\newcommand{\smpy}{M_{\sun}/{\rm yr}}
\newcommand{\cmcmg}{\ensuremath{\mathrm{cm^{2}\,  g^{-1}}  } }
\newcommand{\gcmcm}{\ensuremath{\mathrm{g \, cm^{-2}}  } }

\newcommand{\SigmaAm}{\ensuremath{\Sigma_{\mathrm{Am}}}}
\renewcommand{\fdg}{\ensuremath{f_{\rm dg}}}
\newcommand{\Mgap}{M\sub{gap}}
\newcommand{\Mmig}{M\sub{mig}}
\newcommand{\Mrg}{M\sub{RG}}
\newcommand{\Miso}{M\sub{iso}}
\newcommand{\Mp}{M\sub{p}}
\newcommand{\qp}{q\sub{p}}
\newcommand{\alphat}{\alpha\sub{t}}
\newcommand{\St}{{\rm St}}

\newcommand{\mFid}{\texttt{Fid}\xspace}
\newcommand{\mEff}{\texttt{Vis}\xspace}
\newcommand{\mApz}{\texttt{apz-sgm}\xspace}
\newcommand{\mDG}{\texttt{DG}\xspace}

\defcitealias{Mori2021Evolution-of-th}{M21}

\begin{document}

\titlerunning{Evolution of the disk temperature and protoplanets in magnetized disks}
\authorrunning{Mori et al.}

\title{Long-term evolution of the temperature structure in magnetized protoplanetary disks and its implication for the dichotomy of planetary composition}

\author{
    Shoji Mori\inst{1,2}\thanks{e-mail: mori.s@astr.tohoku.ac.jp}
\and Masanobu Kunitomo\inst{3,4}
\and Masahiro Ogihara\inst{5,6}}

\institute{
Institute for Advanced Study and Department of Astronomy, Tsinghua University, Beijing 100084, China
\and
Astronomical Institute, Tohoku University, 6-3 Aramaki, Aoba-ku, Sendai 980-8578, Japan
\and
Department of Physics, Kurume University, 67 Asahimachi, Kurume, Fukuoka 830-0011, Japan
\and
 Universit\'e C\^ote d'Azur, Observatoire de la C\^ote d'Azur, CNRS, Laboratoire Lagrange, Bd de l'Observatoire, CS 34229, 06304 Nice cedex 4, France
\and
Tsung-Dao Lee Institute, Shanghai Jiao Tong University, 1 Lisuo Road, Shanghai 201210, China
\and
School of Physics and Astronomy, Shanghai Jiao Tong University, 800 Dongchuan Road, Shanghai 200240, China
}

\abstract
{The thermal structure and evolution of protoplanetary disks play a crucial role in planet formation. In addition to stellar irradiation, accretion heating is also thought to significantly affect the disk thermal structure and planet formation processes.}
{
We present the long-term evolution (from the beginning of Class II to disk dissipation) of thermal structures in laminar magnetized disks to investigate where and when accretion heating is a dominant heat source.
In addition, we demonstrate that the difference in the disk structures affects the water content of forming planets.
}{
We considered the mass loss by magnetohydrodynamic (MHD) and photoevaporative disk winds to investigate the influence of wind mass loss on the accretion rate profile. Our model includes the recent understanding of accretion heating, that is, accretion heating in laminar disks is less efficient than that in turbulent disks because the surface is heated at optically thinner altitudes and energy is removed by disk winds.
}{
 We find that accretion heating is weaker than irradiation heating at about 1--10 au even in the early Class II disk, but it can affect the temperature in the inner 1 au region.
     We also find that the magnetohydrodynamic wind mass loss in the inner region can significantly reduce the accretion rate compared with the rate in the outer region, which in turn reduces accretion heating.
     Furthermore, using evolving disk structures, we demonstrate that when accretion heating models are updated, the evolution of protoplanets is affected. In particular, we find that our model produces a dichotomy of the planetary water fraction of 1--10 $M_\oplus$.
}{}


\maketitle
\nolinenumbers

\section{Introduction}
\label{sec:intro}

The temperature structure in protoplanetary disks (PPDs) impacts many aspects of planet formation.
The disk temperature determines the location of the water snowline \citep[e.g.,][]{Hayashi1979Earths-melting-,Oka2011Evolution-of-Sn,Morbidelli2016Fossilized-cond,Chambers2023Making-the-Sola,Wang2025Solving-for-the} and the water content of the forming terrestrial planets \citep[e.g.,][]{Raymond2005Terrestrial-Pla,Mulders2013Why-circumstell,Sato2016On-the-water-de,Ida2019aWater-delivery-,Venturini2020Most-super-Eart,Izidoro2021Formation-of-pl}.
In addition, the temperature affects planetary formation processes, such as the halting of pebble accretion onto protoplanets \citep[i.e., pebble isolation; e.g.,][]{Lambrechts2014Separating-gas-,Lambrechts2014Forming-the-cor,Bitsch2015aThe-growth-of-p}, the mass of objects that formed via the streaming instability \citep[e.g.,][]{Schafer2017Initial-mass-fu,Liu2020Pebble-driven-p}, and the gas-giant migration rate and gap-forming mass \citep[e.g.,][]{Lin1979Tidal-torques-o,Goldreich1979The-excitation-,Tanaka2002Three-Dimension,Paardekooper2010A-torque-formul,Kanagawa2018Radial-Migratio}.

Whereas the temperature profile in the outer region is determined by irradiation heating, the inner profile is determined by accretion heating.
The former is due to stellar irradiation, and the latter mechanism is due to the energy that is released through mass accretion.
In classical disk models, the disk is assumed to be turbulent; thus, turbulent viscosity causes energy dissipation in accretion heating \citep[i.e., viscous heating;][]{Shakura1973Black-holes-in-,Lynden-Bell1974The-evolution-o}.
The heat produced by viscous heating is released around the midplane, which we refer to as midplane heating.
When the disk is optically thick for radiative cooling from the midplane heat, the heat accumulates in the disk, and this may efficiently increase the disk temperature \citep[blanketing effect; see][]{Hubeny1990Vertical-struct,Nakamoto1994Formation-early,Oka2011Evolution-of-Sn,Savvidou2020Influence-of-gr}.

Nevertheless, the presence of strong disk turbulence is doubtful.
Magnetorotational instability \citep[MRI;][]{Balbus1991A-powerful-loca} may generate vigorous turbulence, but it is suppressed by nonideal magnetohydrodynamic (MHD) effects (i.e., Ohmic diffusion, ambipolar diffusion, and the Hall effect) in the weakly ionized gas of PPDs \citep[e.g.,][]{Gammie1996Layered-Accreti,Sano2002aThe-Effect-of-t,Perez-Becker2011Surface-Layer-A,Bai2013aWind-driven-Acc,Bai2013bWind-driven-Acc,Gressel2015Global-Simulati,Bai2017Global-Simulati,Iwasaki2024Dynamics-Near-t}.
Disk observations also suggest the absence of strong turbulence in the disk interior \citep[see the review of][]{Rosotti2023Empirical-const}.

Therefore, instead of turbulent disks, laminar disk models with accretion that is driven by global-scaled magnetic fields have attracted attention.
The stress of magnetic fields that thread the disk removes the angular momentum from the disk through disk winds \citep[e.g.,][]{Blandford1982Hydromagnetic-f,Shibata1986A-magnetodynami,Kudoh1995Mass-Flux-and-T,Suzuki2009Disk-Winds-Driv, Bai2013aWind-driven-Acc,Bai2013bWind-driven-Acc,Bai2016Magneto-thermal}, which is referred to as wind-driven accretion.
When the magnetic flux is strong enough, this mechanism efficiently drives disk accretion and may result in accretion rates that agree with observational data \citep[e.g.,][]{Suzuki2016Evolution-of-pr,Bai2016Towards-a-Globa,Tabone2020Molecule-format,Lesur2021Systematic-desc,Tabone2022Secular-evoluti,Weder2023Population-stud}.

\begin{figure*}[t!]
    \centering
    \includegraphics[width=.9\linewidth]{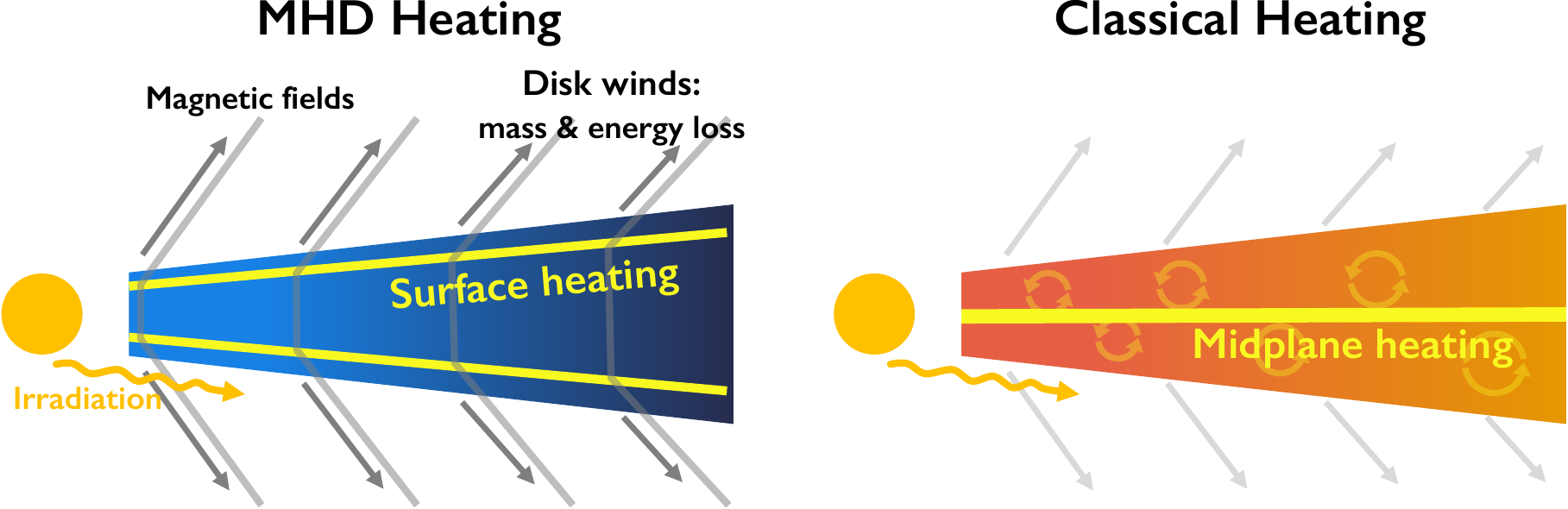}
    \caption{
    Schematic picture of the two heating models shown in this paper: MHD heating, and classical heating.
    The MHD heating model includes the effects where (1) accretion heating occurs at a high altitude, (2) the accretion energy is taken by disk winds, and (3) the accretion rate is reduced by the mass loss of the winds.
    In the classical heating model, accretion heating occurs at the midplane, and the (2)--(3) wind effects are weak.
    Both models include irradiation heating.
             \label{fig:sketch}
    }
\end{figure*}

Accretion heating in magnetic laminar disks is less efficient than the heating in classical turbulent disks (see Fig. \ref{fig:sketch}). Nonideal MHD simulations \citep{Hirose2011Heating-and-Coo,Mori2019Temperature-Str} have shown that the accretion energy that is stored in magnetic fields dissipates away from the midplane.
The heat is released in optically thinner regions, and thus, the cooling is more efficient.
\citet{Mori2021Evolution-of-th} \citepalias[hereafter referred to as][]{Mori2021Evolution-of-th} modeled the surface heating \citep{Mori2019Temperature-Str} and demonstrated that accretion heating can be remarkably inefficient.
\citet{Kondo2023The-Roles-of-Du} updated the \citetalias{Mori2021Evolution-of-th} model by considering dust growth, and the authors showed that it increased the ionization fraction with a higher opacity. This caused the impact of accretion heating to be stronger than was obtained by \citetalias{Mori2021Evolution-of-th}.

Although previous studies have simplified the treatment of the accretion rate (\citetalias{Mori2021Evolution-of-th}; \citealt{Kondo2023The-Roles-of-Du}), this simplification may affect the temperature structure. A uniform distribution of the accretion rate was assumed by neglecting the mass loss.
The strong disk mass loss can affect the accretion rate.
Moreover, the evolution of the accretion rate was assumed to follow the observational relation between the stellar accretion rate and age.
However, the accretion rate evolution of an individual object does not necessarily follow the observational relation obtained from numerous stars.

In this study, we calculate the long-term evolution of the temperature structure in laminar magnetized PPDs, with the mass loss by MHD and photoevaporative winds.
This allows us to evaluate the impact of the simplification made in previous studies. Additionally, whereas previous studies focused on the temperature structure around 1~au and on snowline migration, we pose the more general question where and when accretion heating affects the disk temperature.
We define the region in which accretion heating dominates as the accretion-heated region, where the temperature depends on factors such as the accretion rate, dust opacity, and ionization fraction.
Outside this region, the temperature can be approximated by conventional passively heated disk models.

Furthermore, we investigate the effect of the disk temperature structure on planetary evolution, with a particular focus on the water content.
Recent exoplanet observations have revealed a dichotomy in the volatile composition of close-in planets within the range of 1--10~$M_\oplus$ \citep{Luque2022Density-not-rad,Parviainen2024SPRIGHT:-a-prob,Parc2024From-super-Eart}.
Specifically, according to \citet{Parc2024From-super-Eart}, volatile-rich planets are found above 4.2 $M_\oplus$ around FG-type stars.
\citet{Rogers2025Most-Super-Eart} estimated that the water fraction of super-Earths is smaller than a few percent.
The migration of planets that form beyond the snowline plays a significant role in shaping their compositions \citep[e.g.][]{Venturini2020Most-super-Eart,Izidoro2021Formation-of-pl,Izidoro2022The-Exoplanet-R,Burn2024A-radius-valley}.
We explore how planetary growth and migration shape the dichotomy in planetary water content based on different disk models.
Although processes such as planet photoevaporation \citep[e.g.,][]{Owen2013Kepler-Planets:,Owen2017The-Evaporation,McDonald2019The-Sub-Neptune,Affolter2023Planetary-evolu} and core-powered mass loss \citep[e.g.][]{Ginzburg2018Core-powered-ma,Gupta2019Sculpting-the-v}, which release hydrogen after migration, may also affect the water content, a comprehensive analysis incorporating these effects is left for future study.

In Sect. \ref{sec:method} we describe the method and models we used in the one-dimensional disk evolution simulation.
Section \ref{sec:results} presents the simulation results with a focus on the accretion-heated region. We also perform parameter studies.
In Sect. \ref{sec:discussion} we describe the updates from previous works and discuss the impact of the thermal structure on the evolution process and water fraction of protoplanets.

\section{Method} \label{sec:method}

We designed the disk model to be a wind-driven accretion disk (see Sect. \ref{ssec:basiceq} for the basic equations) with the mass loss (Sect. \ref{ssec:wind-mass-loss}).
Accretion was mainly driven by the removal of the angular momentum by wind stress, and weak radial stress was also considered.

For the thermal structure, we considered two heating models, as depicted in Fig. \ref{fig:sketch}: MHD heating, and classical viscous heating (see Sect. \ref{ssec:Tmid}).
Our MHD heating model considered surface heating depending on the vertical profile of the ionization fraction.
The energy loss and mass loss by the disk winds were also considered.
The latter model, the classical viscous heating model, corresponds to the weak DW case by \citet{Suzuki2016Evolution-of-pr}.
In this model, heating takes place at the midplane. Although energy loss and mass loss through disk winds may occur, these losses are weak.
The difference in the two models highlights the update from the conventional viscous-heating disks.

As an update from \citetalias{Mori2021Evolution-of-th}, we selected simulation parameters based on the median values of plausible disk parameters to derive a representative temperature structure in our model.
This is different from \citetalias{Mori2021Evolution-of-th}, who chose parameters to maximize accretion heating.
For the energy fraction of accretion heating, $\erad$, we adopted a typical value suggested by MHD simulations (see Sects. \ref{ssec:Tmid} and \ref{ssec:setting}). Additionally, we updated the stellar luminosity evolution (see Sect. \ref{ssec:Tirr}).

\subsection{Basic equation} \label{ssec:basiceq}

We solved the one-dimensional advection--diffusion equation for the surface density \citep[e.g.,][]{Lynden-Bell1974The-evolution-o,Suzuki2016Evolution-of-pr,Kunitomo2020Dispersal-of-pr}:
\begin{equation}
\begin{split}
 \pd{\Sigma}{t}  -  \frac{1}{r}\frac{\partial}{\partial r}
  \left[\frac{2}{r\Omega}\left\{\frac{\partial}{\partial r}(r^2 \Sigma
  \overline{\alpha_{r\phi}}c_{\rm s}^2) + r^2 \overline{\alpha_{\phi z}}
  (\rho c_{\rm s}^2)_{\rm mid} \right\}\right] \\
 +  \SigdotDW + \SigdotPE =0\,
\end{split},
  \label{eq:sgmevl}
\end{equation}
where $\Sigma$ is the gas surface density, $t$ is the time, $r$ is the distance from the central star, $\rho$ is the gas density, the subscript mid stands for the midplane, $\SigdotDW$ is the MHD wind mass-loss rate, and $\SigdotPE$ is the photoevaporation rate.
The midplane sound speed $\cs$ is given by the midplane temperature $\Tmid$ as $\cs = \sqrt{ \kB\Tmid /(\mu \amu) }$, where $\kB$ is the Boltzmann constant, $\mu = 2.34$ is the mean molecular weight, and $\amu$ is the atomic mass.
The gas scale height is $H=\cs/\Omega$, where $\Omega=\sqrt{G\Mstar/r^3}$, with $G$ being the gravitational constant and $\Mstar$ being the stellar mass.
The midplane density is given by $\rhomid=\Sigma/(\sqrt{2\pi}H)$.
We neglected disk self-gravity and pressure-gradient forces.
We used the same numerical scheme as \citet{Kunitomo2020Dispersal-of-pr} to solve Eq.\,(\ref{eq:sgmevl}) (e.g., grid spacing, explicit code, and boundary conditions).

The terms $\arp$ and $\apz$ in Eq. (\ref{eq:sgmevl}) describe the radial mass flow due to the radial and vertical angular momentum transport, respectively.
These two parameters model the strength of the radial and vertical stresses as
\begin{align}
  \int \dd z \left(\rho v_r \delta v_{\phi} - \frac{B_rB_{\phi}}{4\pi}\right)
  &\equiv \overline{\alpha_{r\phi}}\, \Sigma  c_{\rm s}^2
  \label{eq:rphistress} , \\
  \left(\rho \delta v_{\phi} v_z - \frac{B_{\phi} B_z}{4\pi}\right)_{\rm w}
 &\equiv \overline{\alpha_{\phi z}}\, \rho_{\rm mid} c_{\rm s}^2 \,,
  \label{eq:phizstress}
\end{align}
where $\bm{B}$ represents the magnetic field, $\bm{v}$ indicates the velocity, and $\delta v_\phi$ is the deviation from Keplerian velocity. The parentheses with the subscript w conduct the division between the values at the top and bottom of the wind base. The choice of these parameter values is described in Sect. \ref{ssec:setting}.

\subsection{Temperature structure}\label{ssec:Tmid}

As shown in Fig. \ref{fig:sketch}, we considered the MHD heating model and the classical heating model.
In both models, the midplane temperature $\Tmid$ is given by
\begin{equation}\label{eq:tmid-sum}
        \Tmid= \pr{ \Tacc^4+\Tirr^4 }^{1/4},
\end{equation}
where $\Tacc$ and $\Tirr$ are the midplane temperatures solely due to accretion heating and irradiation heating, respectively.
The contribution by accretion heating, $\Tacc$, depends on the heating model. The treatment of $\Tirr$ is common between the two heating models (see Sect. \ref{ssec:Tirr}).

\subsubsection{Accretion heating}\label{sec:Tacc}

For the MHD heating and classical models, we determined $\Tacc$ by following \citetalias{Mori2021Evolution-of-th} (see their Appendix A),
\begin{equation}\label{eq:Tmid}
        T_{\rm acc} = \left[ \pr{ \frac{3 \Frad }{8 \sigma} }
        \pr{   \kappa \sgmheat   +  \frac{1}{\sqrt{3}} }  \right] ^{1/4} ,
\end{equation}
where
$\Frad$ is the energy flux that escaped as radiation from the disk,
$\sigma$ is the Stefan--Boltzmann constant,
$\sgmheat$ is the column density above the heating layer,
$\kappa$ is the Rosseland mean opacity for the radiation from the disk,
and the term in the optically thin region was neglected.
We adopted a simple opacity model as a function of $T_{\rm mid}$, similar to \citet{Kunitomo2020Dispersal-of-pr},
\begin{equation}
\label{eq:kappa}
        \kappa = \kappa_0 \,\left [1 + \exp\pf{T_{\rm mid} - T_{\rm sub} }{\Delta T/2 } \right]^{-1} \!\! \min\left[ 1, \pf{T_{\rm mid}}{170\,\K}^2 \right] ,
\end{equation}
where the reference Rosseland mean opacity $\kappa_0 $ was set to $5\,  \cmcmg$, $T_{\rm sub} = 1500\,\K$, and $\Delta T = 150\,\K$.

We varied $\sgmheat$ depending on the heating model.
In the classical heating model, the heating takes place at the midplane; thus, we set $\sgmheat$ to be $\Sigma / 2$.
In the MHD heating model, we gave $\sgmheat$ as the mass column density $\Sigma_{\rm Am}$, where the ambipolar Elsasser number ${\rm Am}$ is a critical value \citepalias[see][for details]{Mori2021Evolution-of-th}.
We took the critical value to be 0.3 as in \citetalias{Mori2021Evolution-of-th}.
This approach is based on the simulation results, according to which the accretion heating occurs at the thin layer where ${\rm Am} \sim 1$ (see Fig. 1 of \citet{Mori2019Temperature-Str}).
To obtain the vertical ${\rm Am}$ distribution for each radius, we calculated the ionization fraction along the vertical direction by assuming hydrostatic equilibrium and then calculated ${\rm Am}$, as in \citetalias{Mori2021Evolution-of-th} \citep[see also][]{Mori2016Electron-Heatin}.

We parameterized the total heating rate $\Frad$ by the fraction $\erad$ of the dissipated energy in the generated energy, as in \citet{Suzuki2016Evolution-of-pr},
\begin{equation}\label{eq:Frad}
         \Frad = \erad \pr{ \Grp+\Gpz },
\end{equation}
where $ \Grp$ and $\Gpz $ are the liberated energy due to the $r\phi$- and $\phi z$-component stresses, respectively.
These are calculated by
\begin{align}
        \Grp &= \frac{3}{2}   \Omega \Sigma \overline{\alpha_{r \phi}} c_{\mathrm{s}}^{2} , \\
        \Gpz &= r \Omega \overline{\alpha_{\phi z}} \rhomid c_{\mathrm{s}}^2  .
\end{align}

\subsubsection{Irradiation heating}
\label{ssec:Tirr}

We considered two regimes for the midplane temperature due to irradiation, depending on the optical depth of the stellar direct radiation.
In the optically thick region, according to \citet{Kusaka1970Growth-of-Solid} and \citet{Chiang1997Spectral-Energy},
\begin{equation}\label{eq:Tirr_1}
        T\sub{irr,thick} = 110\,\K
 \left( \frac{r}{\au} \right)^{-3/7}
 \left( \frac{ \Lstar }{ \Lsun } \right)^{2/7}
 \left( \frac{ \Mstar }{ \Msun } \right)^{-1/7}
 \,,
\end{equation}
where $\Lstar$ is the stellar luminosity, and half of the stellar hemisphere was assumed to illuminate the disk surface.
When the disk is optically thin for the stellar light through the midplane, we wrote \citep{Hayashi1981Structure-of-th}
\begin{equation}\label{eq:Tirr_2}
        T\sub{irr,thin} = 280\,\K \left( \frac{r}{\au} \right)^{-1/2} \left( \frac{ \Lstar }{ \Lsun } \right)^{1/4} \,.
\end{equation}
We switched these regimes using the irradiation optical depth ($\tau\sub{*}(r)=\int_0^r\kappa\sub{*}\rho\sub{mid}\d r'$) along the midplane, where $\kappa\sub{*}$ is the Planck mean opacity for the stellar light,
\begin{equation}\label{eq:Tirr}
        \Tirr^4 = T\sub{irr,thin}^4 +  T\sub{irr,thick}^4 \exp \pr{ -\tau\sub{P} } \,.
\end{equation}
We set $\kappa\sub{*}=10\,\rm cm^{2}\,g^{-1}$, but the results are insensitive to this value.
In addition, no gradual temperature transition \citep[e.g.,][]{Oka2011Evolution-of-Sn,Okuzumi2022A-global-two-la} was considered.
However, the intermediate state would appear only on a short timescale because the surface density quickly drops because of photoevaporation.

We used a more realistic evolution model of stellar luminosity $\Lstar$ than that of \citetalias{Mori2021Evolution-of-th}, where the $\Lstar$ evolution model was the same as that by \citet{Feiden2016Magnetic-inhibi}.
We also adopted the model by \citet{Feiden2016Magnetic-inhibi}, but with a modification: Because the initial luminosity in their model is higher than expected from those of star formation models, we shifted $t=0$ to the timing when an evolutionary track on the Hertzsprung--Russell diagram crosses the birthline reported by \citet[][see also Fig. 1 of \citealt{Kunitomo2021Photoevaporativ}]{Stahler2004The-Formation-o}.
As shown below, the luminosity in the early phase ($t \lesssim 0.5$\,Myr) is lower than that used by \citetalias{Mori2021Evolution-of-th}, which leads to the earlier arrival of the snowline at 1~au ($\sim$ 0.2~Myr).

\subsection{MHD and photoevaporative winds}\label{ssec:wind-mass-loss}

For the mass loss by MHD disk wind, we followed the approach by \citet{Suzuki2016Evolution-of-pr}.
The mass-loss rate was parameterized as $\Cw \equiv \SigdotDW/(\rhomid \cs)$, which is the mass flux normalized by $\rhomid \cs$.
In this model, the mass-loss parameter $\Cw$ is given by a constant $\cwo$ unless an energy constraint of the disk winds is violated.
The energy constraint is given from the primary energy balance \citep{Suzuki2016Evolution-of-pr}:
\begin{equation}\label{eq:suzuki-enebalance}
        \SigdotDW \pr{E_{\rm w} + r^2\Omega^2/2 } + \Frad \sim \Grp+\Gpz,
\end{equation}
where $E_{\rm w}$ is the specific energy density of the magnetic disk wind.
To launch the disk wind, the wind energy $E_{\rm w}$ must be positive.
Using Eq. (\ref{eq:Frad}), we have the constraint
\begin{equation}
\label{eq:cwe}
        \Cw \gtrsim \dfrac{2 \pr{1 - \erad}  \pr{\Grp+\Gpz} }{\rhomid \cs r^2\Omega^2}  \equiv \cwe,
\end{equation}
where $\cwe$ is the mass-loss parameter required from the energy constraint.
We assumed that the energy balance due to irradiation heating was independently satisfied.
The irradiation heating rate and its cooling rate were therefore not seen.
For $\cwo$, we adopted the same value as \citet{Suzuki2016Evolution-of-pr}.
Although the mass-loss rate may vary with different simulations, we adopted $\cwo = 10^{-5}$ as used by \citet{Suzuki2016Evolution-of-pr}.
Thus, we give the mass-loss rate of the MHD disk wind as
\begin{equation}
\label{eq:sigdotdw}
        \SigdotDW = \max\pr{ \cwo, \cwe} \rhomid \cs .
\end{equation}
Although $\Cw$ depends on the magnetic flux that threads the disk \citep[e.g.,][]{Bai2013aWind-driven-Acc,Bai2013bWind-driven-Acc,Bai2017Global-Simulati,Lesur2021Systematic-desc}, we did not consider this dependence.

Furthermore, we considered mass loss through photoevaporation. Young protostars are active and emit intense extreme-ultraviolet (EUV), far-ultraviolet (FUV), and X-rays.
When it is exposed to these energetic rays, the gas in the upper disk becomes so hot that it exceeds the escape velocity from the stellar gravitational potential and flows out of the disk.
For the photoevaporative mass-loss rate $\SigdotPE$, we followed \citet[][see their Sect. 2.4]{Kunitomo2020Dispersal-of-pr}: We considered the X-ray ($\SigdotX$) and EUV ($\SigdotEUV$) photoevaporation rates in the literature \citep{Alexander2006aPhotoevaporatio,Alexander2007Dust-dynamics-d,Owen2012On-the-theory-o} with an X-ray luminosity of $10^{30}\,\rm erg/s$ and an EUV luminosity of $10^{41}\,\rm s^{-1}$. Their sum is the mass-loss rate $\SigdotPE$,
\begin{equation}
        \SigdotPE = \SigdotX+\SigdotEUV.
\end{equation}
Because the inner disk accretes rapidly onto the star and the outer disk is directly irradiated, we switched $\SigdotPE$ as in \citet[][]{Kunitomo2020Dispersal-of-pr} when a gap opened.
We note that although $\SigdotPE$ is still actively studied and uncertain \citep[see, e.g., discussions in Sect. 5.4 of ][]{Kunitomo2021Photoevaporativ}  \citep[see also][]{Wang2017Hydrodynamic-Ph, Picogna2019The-dispersal-o, Komaki2021Radiation-Hydro, Nakatani2024Broadening-the-}, this does not affect the conclusion of this study.

\subsection{Simulation settings and parameter choice} \label{ssec:setting}

\newcommand{\mycolhead}[1]{\colhead{\hspace{.0cm}#1}\hspace{.0cm}}
\begin{table}
    \caption{Common parameter values.\label{tab:input}}
    \centering
    \begin{tabular}{ll}
    \hline\hline
    Parameter & Value \\
    \hline
    Radius of the inner boundary, $\rin$ & 0.01 au\\
    Radius of the outer boundary & $10^4$ au\\
    Viscosity parameter, $\arp$ & $10^{-4}$\\
    Initial disk mass & $0.2\,M_\star$\\
    Initial characteristic radius, $r_{\rm c}$ & 100\,\au\\
    Slope of initial surface density, $p$ & $-1$\\
    Reference opacity, $\kappa_0$ & 5 \cmcmg \\
    \hline
    \end{tabular}
\end{table}

\renewcommand{\mycolhead}[1]{\colhead{\hspace{.6cm}#1}\hspace{.6cm}}

\begin{table}
    \caption{Disk models considered in Sect. \ref{sec:results}. \label{tab:models}}
    \centering
    \begin{tabular}{cccccc}
    \hline\hline
    Name & $\sgmheat$ & $\erad$ & $\apz$  & $f_{\rm dg}$ \\
    \hline
    \mFid      & $\SigmaAm$ & 0.1   & const          & 0.01 \\
    \mEff      & $\Sigma/2$ & 0.9   & const        & N/A \\
    \texttt{eps001} & $\SigmaAm$ & 0.01  & const          & 0.01 \\
    \texttt{eps1}   & $\SigmaAm$ & 1.0   & const          & 0.01 \\
    \mDG       & $\SigmaAm$ & 0.1   & const         & 0.001 \\
    \mApz      & $\SigmaAm$ & 0.1   & $\Sigma$-dep     & 0.01 \\
    \hline
    \end{tabular}
    \tablefoot{
    The column density for the accretion heating, $\sgmheat$, was either taken at the midplane ($\Sigma/2$) or at the altitude where ${\rm Am}=0.3$ ($\SigmaAm$).
    The model of $\apz$ is either a constant value of $10^{-3}$ (const) or depends on $\Sigma$ ($\Sigma$-dep).
    The dust-to-gas ratio $\fdg$ was used only for the ionization fraction calculation, while the Rosseland-mean opacity $\kappa$ in the temperature calculation is given by Eq. (\ref{eq:kappa}).
    }
\end{table}

In Table \ref{tab:input} we summarize the values for the common parameters we used throughout.
Table \ref{tab:models} presents the parameter set in the fiducial model (\mFid) and summarizes the other models we used.
For all runs, the calculation was performed until $t = 10$ Myr, which is sufficiently longer than the typical disk lifetime.

For the initial condition of the surface density, we give $\Sigma$ in the form $\Sigma \propto (r/{\rm au})^{p} \exp(- r/r_{\rm c})$, where the characteristic radius $r_{\rm c} $ was set to 100\,au and $p$ was set to $-1$. The initial disk mass was set to $0.2\,M_{\star}$ in the fiducial setting.

The parameters on the accretion stress ($\apz$ and $\arp$; Eqs. (\ref{eq:rphistress}) and (\ref{eq:phizstress})) are given to express the case in which wind stress dominates turbulent stress.
In the model \mFid, we set $\apz$ to be a constant ($\apz = 10^{-3}$) and $\arp$ to be $\arp=10^{-4}$.

In reality, the $\apz$ value is related to the magnetic flux transport in the disk, which is still highly uncertain.
\citet{Suzuki2016Evolution-of-pr} considered two cases corresponding to (relatively) efficient and inefficient flux transports.
In the former, the relative importance of the magnetic field to the gas pressure may be kept, which then leads to a constant $\apz$.
This corresponds to our fiducial case.
In the latter, the surface density decreases without varying the magnetic flux.
The magnetic field strength and $\apz$ become larger as the surface density decreases.
 We also considered the latter case in Sect. \ref{sec:apz-effect}.

According to \citetalias{Mori2021Evolution-of-th} (see their Appendix B), $\erad$\footnote{\citetalias{Mori2021Evolution-of-th} used $\fheat \equiv \Frad / \Gpz$ instead of $\erad$. Nevertheless, in their case of $\apz \gtrsim \arp$, $\fheat$ takes almost the same value.} mainly depends on the polarity of the magnetic field with ranges of $\sim$ 10$^{-2}$--10$^{0}$ and $\sim$ 10$^{-3}$--10$^{-1}$ for the aligned and anti-aligned magnetic fields to the rotation axis, respectively. We adopted 0.1 as a fiducial value, which is in the range of $\erad$ above. We investigate the impact of varying $\erad$ in Sect. \ref{ssec:fheat-effect}.
We also set $\cwo$ to be 10$^{-5}$, following \citet{Suzuki2016Evolution-of-pr}.

For the dust parameters, we adopted a dust-to-gas ratio $\fdg$ of 0.01 and a reference opacity $\kappa_0$ of $5$ cm$^{2}$ g$^{-1}$.
In Sect. \ref{ssec:dust-effect} we also examine the dust model motivated by \citet{Kondo2023The-Roles-of-Du}, where the dust-to-gas ratio is reduced to 0.001 and the opacity remains unchanged.

\section{Results} \label{sec:results}

We first show the simulation results of the fiducial parameter set (Sect. \ref{ssec:res-fid}) and then the dependence on the model parameters (Sect. \ref{ssec:res-par}). In addition, we demonstrate the effect of the temperature model on planet formation (Sect. \ref{ssec:infl-plf}).

\subsection{Fiducial model}\label{ssec:res-fid}

We present our fiducial model (\mFid) by comparing it with a previous model (\mEff).
\mFid was designed for wind-driven accretion disks with MHD heating: the lower heat conversion fraction is $\erad = 0.1$, with surface heating where $\sgmheat = \Sigma_{\rm Am}$.
\mEff contains classical heating: the higher conversion fraction is $\erad = 0.9$ with midplane heating, $\sgmheat = \Sigma/2$.
In both models, accretion is mainly driven by wind stress.

\newcommand{\figtitle}{
     \hspace{0.09\linewidth} \textsf{\bfseries MHD heating (\mFid)} \hspace{.25\linewidth}  \textsf{\bfseries Classical viscous heating (\mEff)} \\ \vspace{1.3mm}
}
\newcommand{\figtitlen}[2]{
      \hspace{#2\linewidth} \textsf{\bfseries MHD heating (\mFid)} \hspace{#1\linewidth}  \textsf{\bfseries Classical viscous heating (\mEff)} \\ \vspace{1.3mm}
}
\begin{figure*}[t!]
    \centering \figtitle
    \includegraphics[width=.45\linewidth]{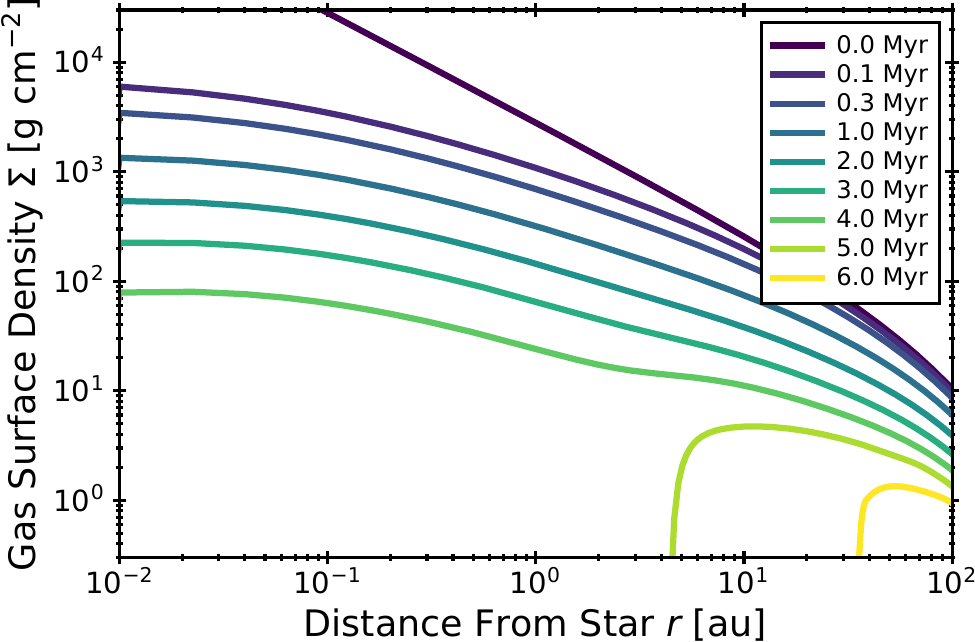}
        \includegraphics[width=.45\linewidth]{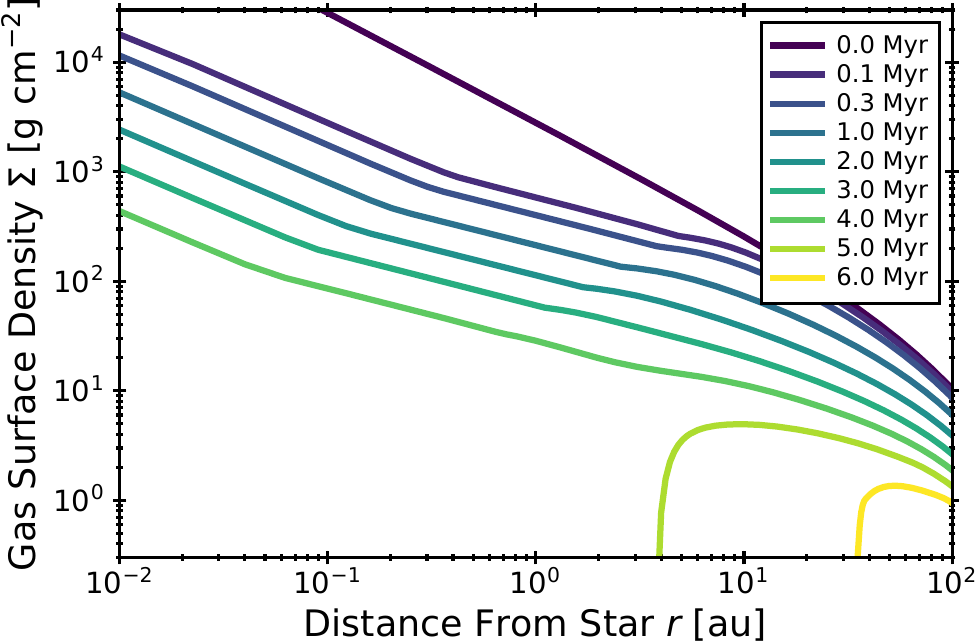}\\ \vspace{3mm}
    \includegraphics[width=.45\linewidth]{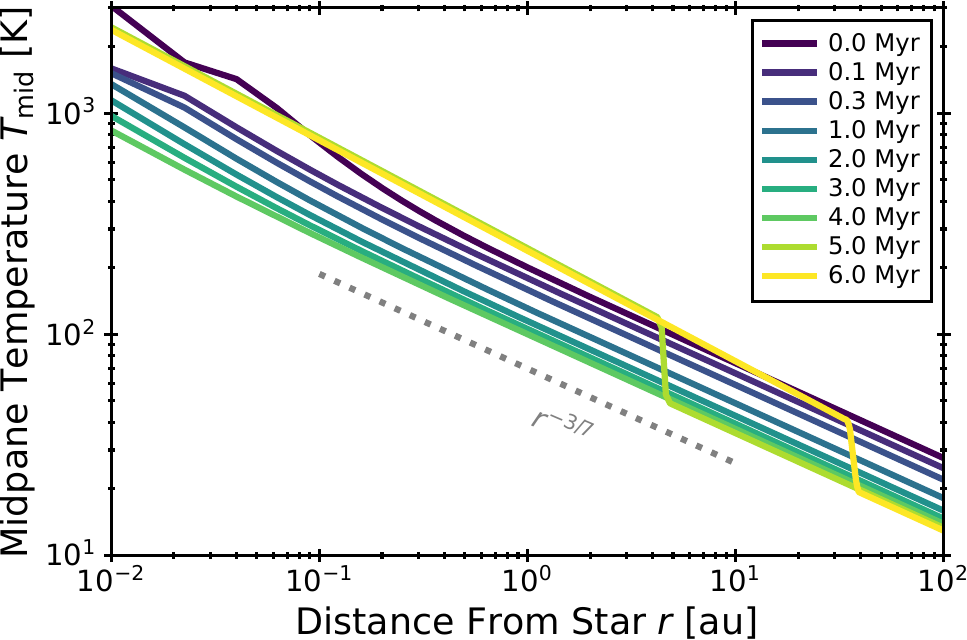}
        \includegraphics[width=.45\linewidth]{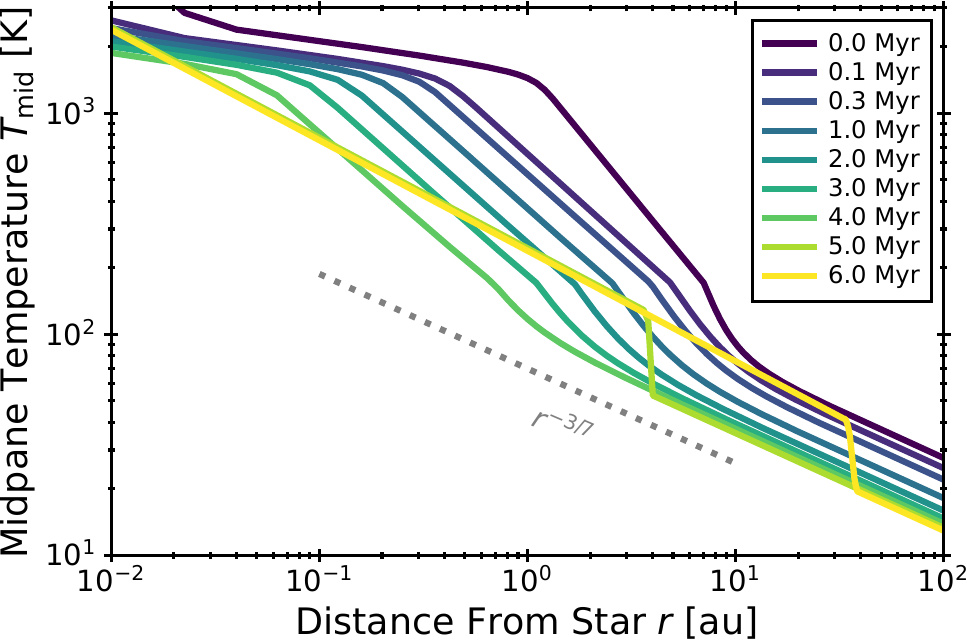}
    \caption{
         Various snapshots of the radial profiles of the gas surface density (top) and midplane temperature (bottom) for the models \mFid (left) and \texttt{Vis} (right).
    \label{fig:prof_Sigma_T_fid}
    }
\end{figure*}

For the basic evolution of the profiles, we show the evolution of the surface density profile in the top panel of Fig. \ref{fig:prof_Sigma_T_fid}.
The basic evolution in \mFid is consistent with that in \mEff, which is similar to \citet{Kunitomo2020Dispersal-of-pr}.
The surface density decreases with time as a result of disk accretion and mass loss. Although the initial surface density profile is given by a radial power of $r^{-1}$ for the inner region, the surface density becomes shallower within the first few hundred years.
We confirmed that in the quasi-steady state, the net mass gains by the disk accretion and mass loss by the disk wind are almost balanced.
At $t\sim$ 5 Myr, disk dissipation begins immediately from the inner region because of photoevaporation.
The disk lifetime is consistent with observational suggestions \citep[$\sim$ 3--10 Myr; e.g.,][]{Williams2011Protoplanetary-}.
Although the overall disk evolution is similar to that of \mEff, the surface density profile within 10 au is affected by the difference in the sound speed caused by temperature variations.

\begin{figure*}[t!]
    \centering \figtitlen{.3}{0.04}
    \includegraphics[width=.48\linewidth]{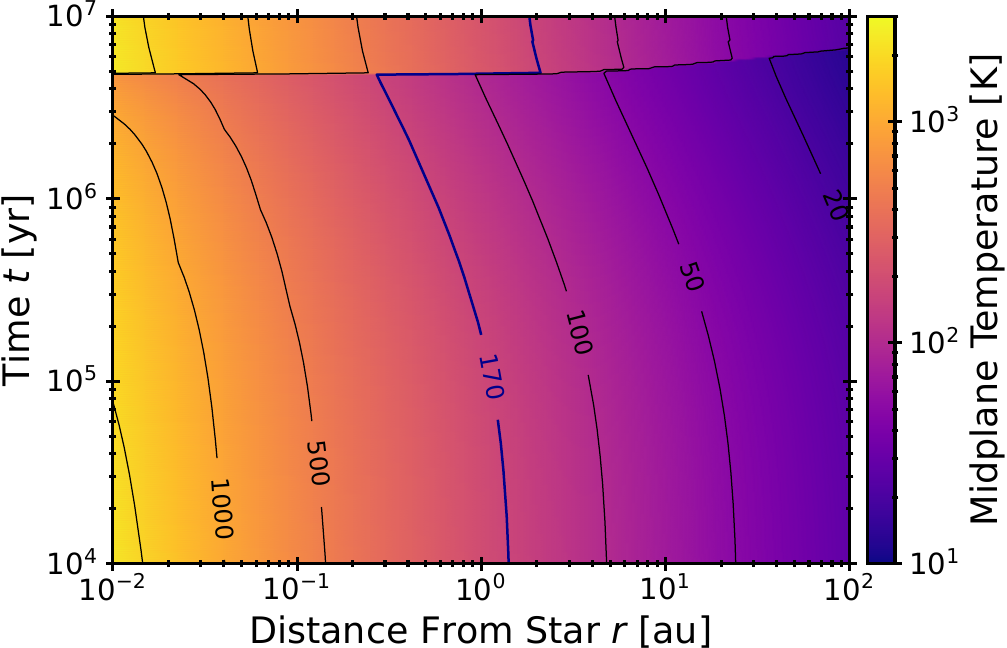}
        \includegraphics[width=.48\linewidth]{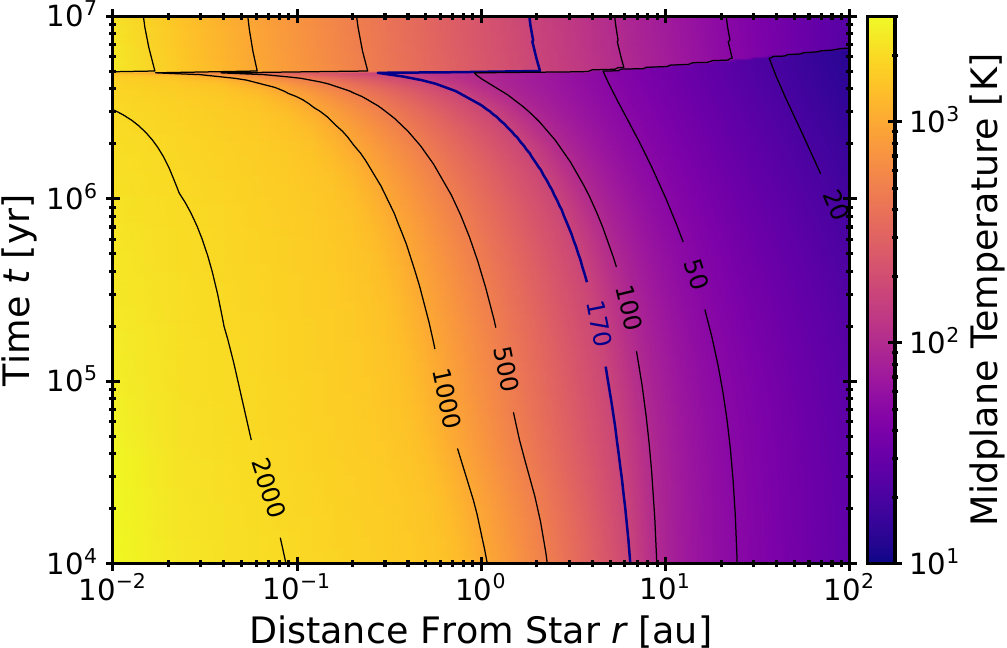}\\
    \includegraphics[width=.48\linewidth]{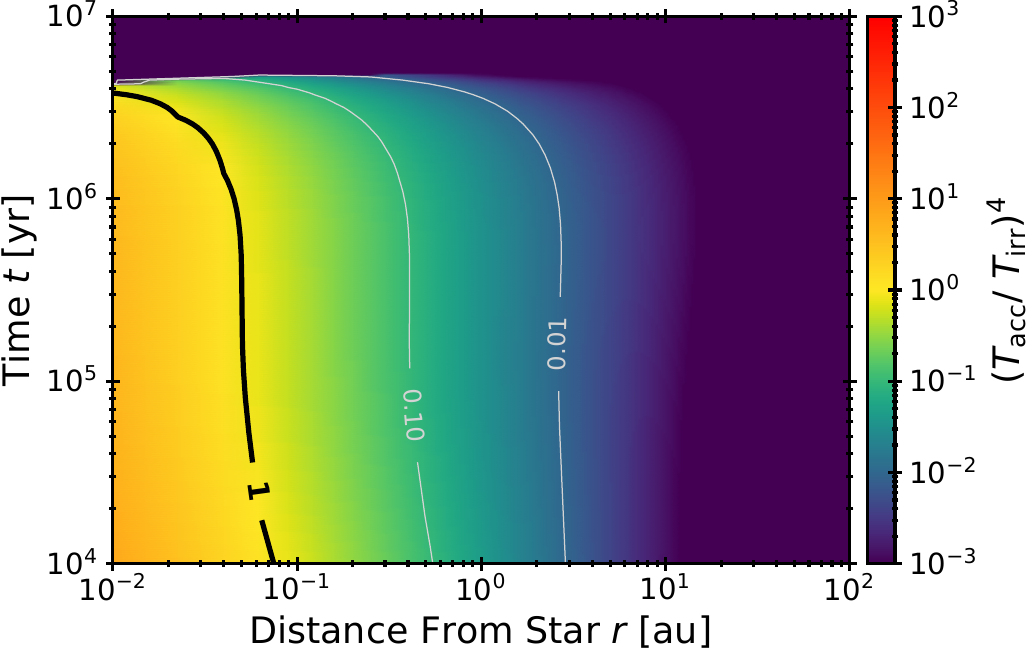}
        \includegraphics[width=.48\linewidth]{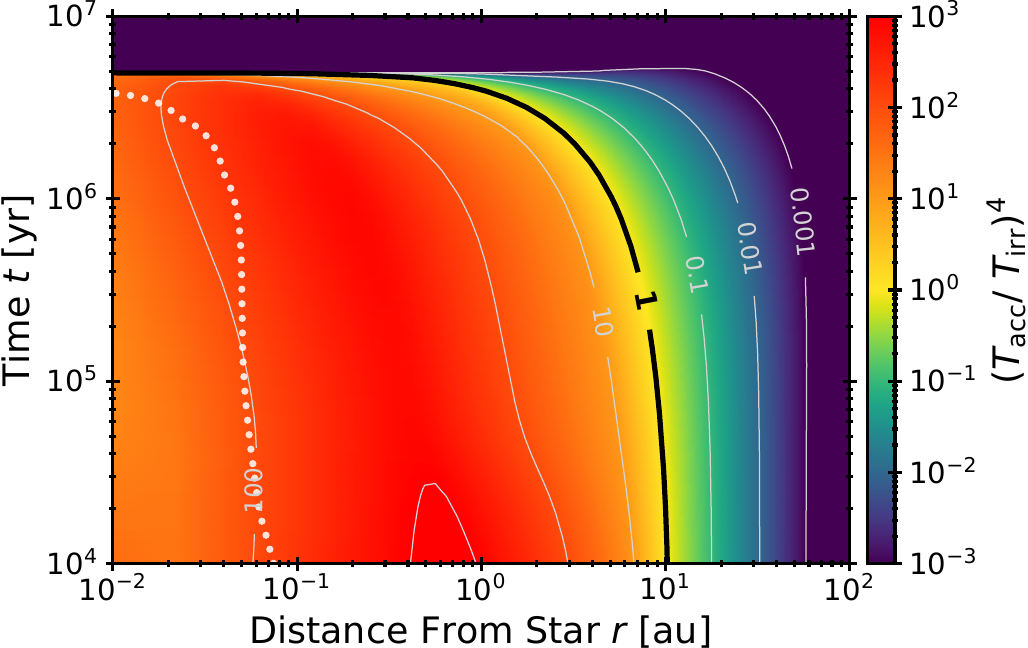}
         \caption{
         Top: Space--time distribution of the midplane temperature.
         The location of the snowline ($\Tmid = 170 \K$) is denoted by the thicker solid line.
          Bottom: Space--time distribution of the fourth power of the temperature ratio $(\Tacc/\Tirr)^4$ between accretion heating and irradiation heating, indicating the contribution of accretion heating to the disk temperature.
          The dotted white line shows the boundary of the accretion-heated region (i.e., $(\Tacc/\Tirr)^4 = 1$) in \mFid for reference.
    \label{fig:Tmid_prof_rt_fid} \label{fig:Tacc_fid}
    }
\end{figure*}

The midplane temperature also evolves with time, as shown in the lower panels of Fig. \ref{fig:prof_Sigma_T_fid}.
For \mFid, the evolution of the temperature profile is mainly driven by the evolution of the stellar luminosity because irradiation heating mainly dominates the midplane temperature, as shown in detail below.
The slope is nearly constant, except during the early stages of disk evolution ($\lesssim 0.1$ Myr), and it is $r^{-3/7}$.
When the surface density in the inner region quickly drops through the photoevaporation,
the direct stellar irradiation becomes able to reach the cavity after $\sim$5\,Myr.
The temperature therefore suddenly increases and transitions at the cavity edge.

Model \mFid shows cooler disk temperatures than \mEff.
This is also shown in the space--time temperature distribution (upper panels of Fig. \ref{fig:Tmid_prof_rt_fid}).
In model \mFid, the iso-temperature contours at higher than 100 K are located at radii that lie more inward.
This is also the case for the snowline, at which $\Tmid = 170$ K.
The snowline in model \mFid reaches 1 au at 0.2 Myr, while that in \mEff reaches 1 au at 4 Myr.
The early arrival of the snowline is consistent with \citetalias{Mori2021Evolution-of-th}.

Next, we show where and when accretion heating is dominant.
We calculated the ratio $(T_{\rm acc}/ T_{\rm irr} )^4$, which represents the contribution to the midplane temperature, since $\Tmid = \Tirr (1 + (T_{\rm acc} / T_{\rm irr})^4 )^{1/4}$ (Eq. (\ref{eq:tmid-sum})).
In particular, the region in which the quantity is larger than unity is the accretion-heated region.
Figure \ref{fig:Tmid_prof_rt_fid} shows $(T_{\rm acc} / T_{\rm irr})^4$ in the space--time diagram.
The accretion-heated region in model \mFid is observed only inside 0.1 au, whereas the region is very wide (within 10 au) in model \mEff.
Even at $r \sim$ 0.3 au, as $(T_{\rm acc}/ T_{\rm irr} )^4 \sim 0.1$, the deviation from the irradiation temperature is only about 3\%.
Therefore, we expect that the accretion heating operates in the limited inner region for the entire disk evolution.

\begin{figure*}[t!]
    \centering
    \figtitle
    \includegraphics[width=.48\linewidth]{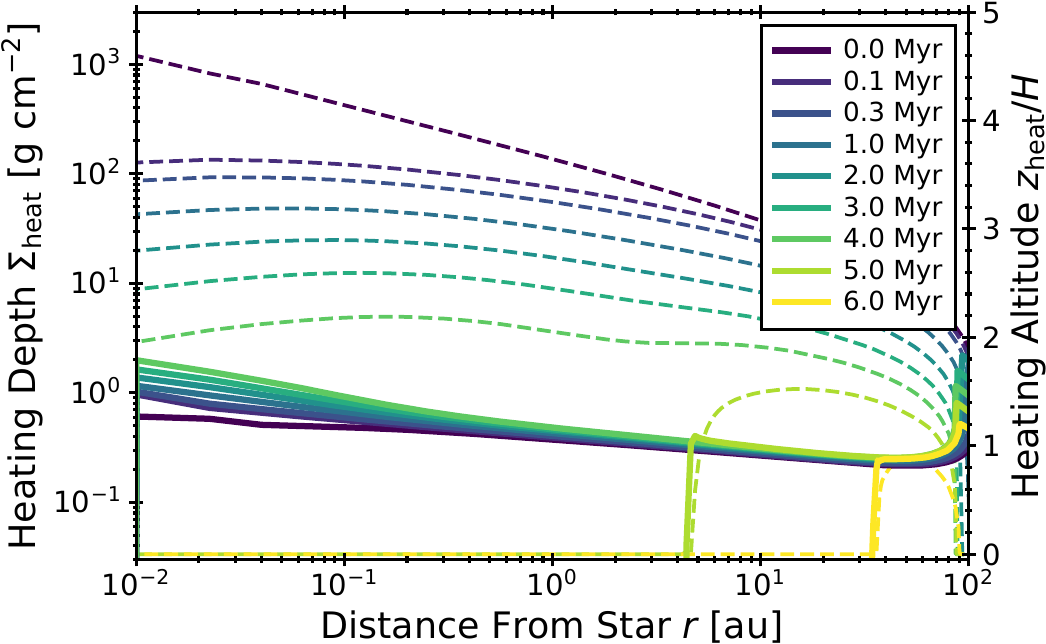}
    \includegraphics[width=.48\linewidth]{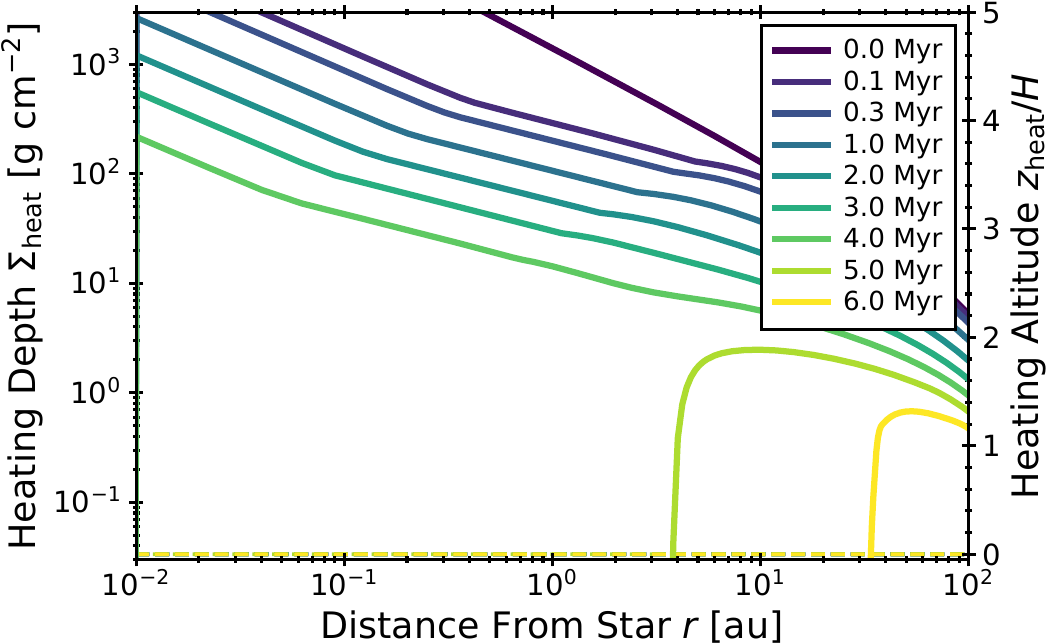}
    \caption{
       Radial profiles of the heating altitude $z_{\rm heat}$ (dashed; right y-axis) and the column density at that altitude, $\sgmheat$ (solid; left y-axis), at various times for models \mFid (left panel) and \mEff (right panel).
    We set $\sgmheat$ to zero when $\Sigma$ is zero.
    \label{fig:Sigmaheat_fid}
    }
\end{figure*}

\begin{figure*}[t!]
    \centering
    \figtitle
    \includegraphics[width=.45\linewidth]{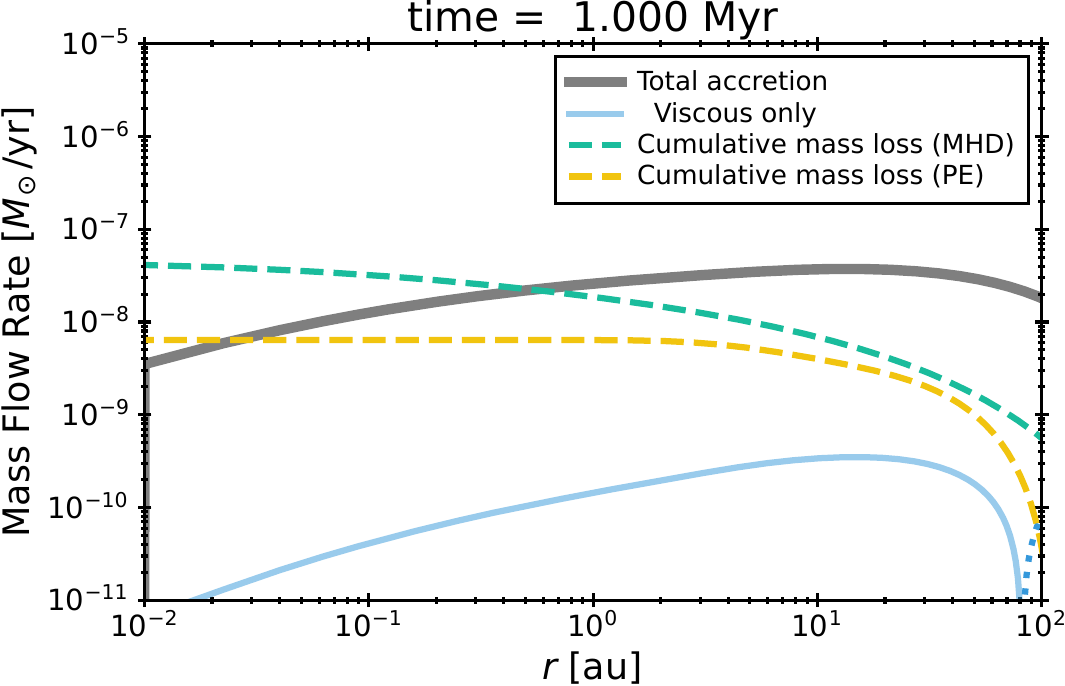}
    \includegraphics[width=.45\linewidth]{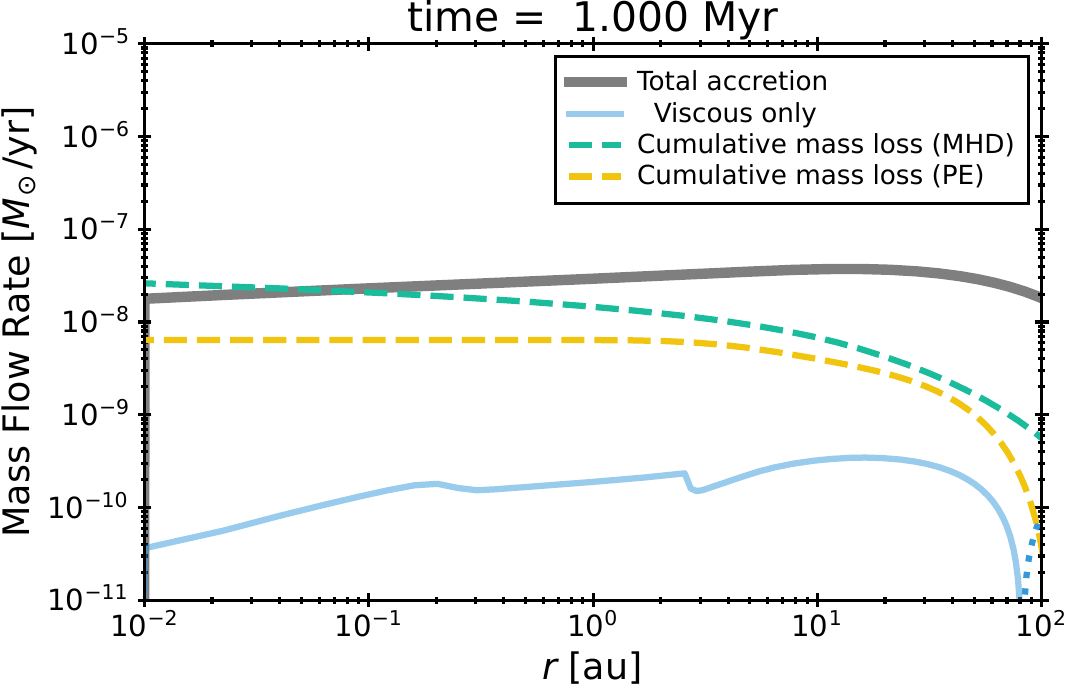}\\
    \includegraphics[width=.45\linewidth]{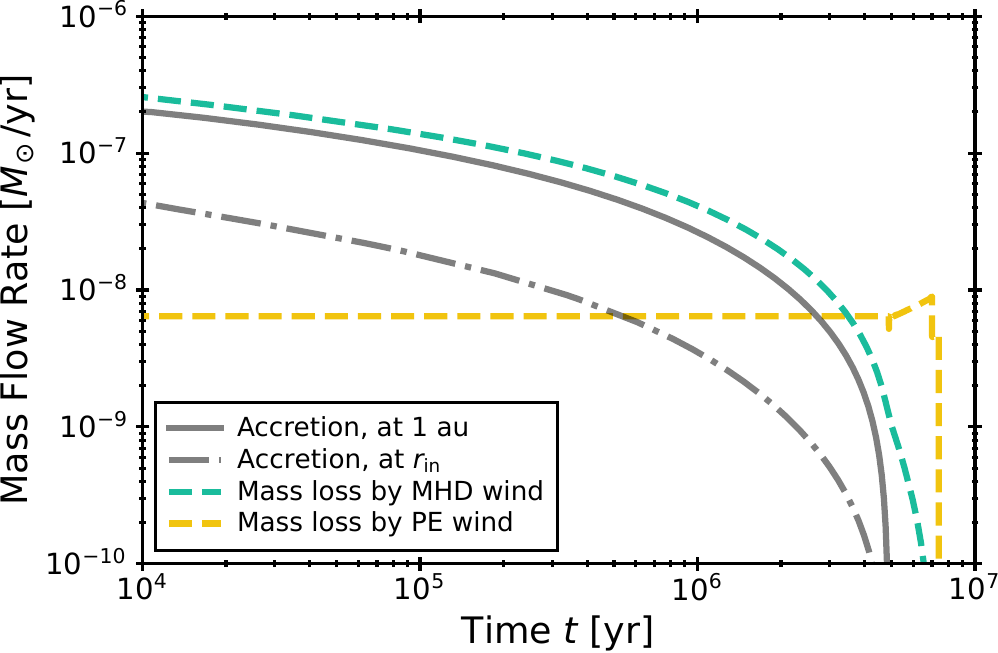}
    \includegraphics[width=.45\linewidth]{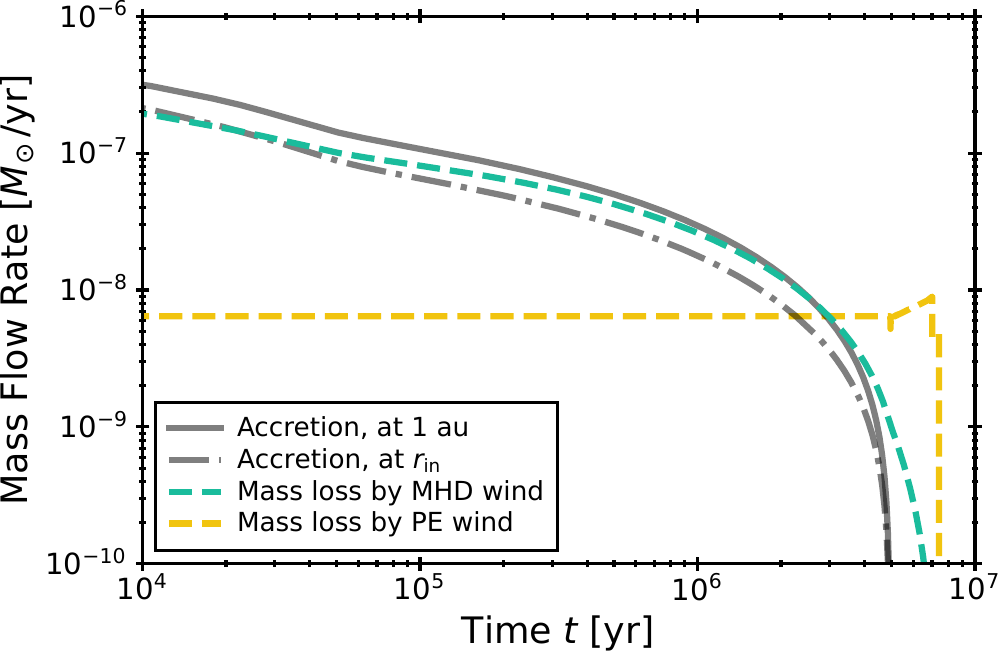}
    \caption{
        Top: Radial profiles of the accretion rate (gray) and the cumulative mass-loss rate due to MHD (dashed green) and photoevaporative winds (dashed yellow) at $t =$ 1 Myr after disk formation. The accretion rate solely due to viscous stress is indicated by blue lines. Model \mFid is shown in the left panels, and \mEff is shown in the right panels.
    Bottom: Time evolutions of the accretion rate measured at $r = \rin$ ($=0.01$\,au; solid gray) and $r = $ 1 au (dashed gray). The total mass-loss rates by the MHD (dashed green lines) and photoevaporative winds (solid yellow) are also shown.
    \label{fig:Mdot_fid}
    }
\end{figure*}

To understand why accretion heating becomes subdominant, we began by analyzing $\sgmheat$.
We plot the distribution of the heating altitude $z_{\rm heat}$ and $\sgmheat$ at various times in Fig. \ref{fig:Sigmaheat_fid}.
As shown in the right panel, $\sgmheat$ in \mEff is $\Sigma /2$, making $\sgmheat$ usually higher than 10 \gcmcm. In contrast, $\sgmheat$ in \mFid, which is $\SigmaAm$, remains $\lesssim$ 1 \gcmcm, at which the heating altitude is $\sim$ 2--3 $H$.
Thus, the difference in $\sgmheat$ can be two orders of magnitude.
In addition, $\erad$ differs by a factor of nine.
These differences significantly suppress the temperature due to accretion heating, as in \citetalias{Mori2021Evolution-of-th}.

Furthermore, because we considered the influence of mass loss on the accretion rate, we analyzed the accretion rate distribution.
The accretion rates induced by the $r\phi$- (e.g., viscous accretion) and $\phi z$-components (wind-driven accretion) stresses are given by
\begin{align}
    \dot{M}_{{\rm acc}, r\phi} &= \frac{4 \pi}{r \Omega} \frac{\partial}{\partial r}\left(r^2 \Sigma \overline{\alpha_{r \phi}} c_{\mathrm{s}}^2\right) , \\
    \dot{M}_{{\rm acc}, z\phi} &= \frac{4 \pi}{\Omega} r \apz \left(\rho c_{\mathrm{s}}^2\right)_{\mathrm{mid}}\,,
\end{align}
respectively \citep[see][]{Suzuki2016Evolution-of-pr}.
We also plot the cumulative mass-loss rate, which is defined as
\begin{equation}
    \dot{M}_{\rm loss}(r)=2 \pi \int_r^{r_{\mathrm{out}}} r \left( \dot{\Sigma}_{\mathrm{MDW}} +
    \dot{\Sigma}_{\mathrm{PEW}} \right) \d r  \,.
\end{equation}

The top left panel of Fig. \ref{fig:Mdot_fid} shows the radial distribution ($t = 1$ Myr) of the accretion rate in \mFid.
The accretion is driven by wind stress (i.e., $\dot{M}_{\rm acc} \sim \dot{M}_{{\rm acc}, z\phi}$) as the viscous accretion rate is much lower than $\dot{M}_{\rm acc}$. $\dot{M}_{\rm acc}$ decreases from 10 au inward.
At $t =$1 Myr, the accretion rate at $r=\rin$ is $3 \times 10^{-9} \smpy$, while the accretion rate at 10 au is $4 \times 10^{-8} \smpy$.
The difference is due to the wind mass loss inside 1 au: 70\% of the mass-loss rate over the whole disk region is responsible for that within 1 au.
In contrast, \mEff shows a weaker mass loss and higher accretion rates.
This is because the mass-loss parameter $\cw$ is constrained by $\cwe$ because of the higher $\erad$ (see Sect. \ref{ssec:wind-mass-loss}).
In other words, the accretion rate is influenced by the heat production efficiency in the disks.

The lower left panel of Fig. \ref{fig:Mdot_fid} shows the time evolution of the accretion rate at $\rin$ and at 1 au over the disk lifetime.
The difference in the accretion rates is almost one order of magnitude until disk dissipation.
This clearly illustrates that wind mass loss reduces the accretion rate in the inner region.

From another perspective, this effect suggests that disk models based on the observed stellar accretion rate may underestimate the actual accretion heating within the disk. However, because of the energy removal by the disk winds and the inefficiency of surface heating, the influence of accretion heating is still considered negligible.

\subsection{Parameter dependence}\label{ssec:res-par}

\subsubsection{Influence of $\erad$}
\label{ssec:fheat-effect}
\begin{figure*}
         \centering
              \hspace{0.05\linewidth}
              \textsf{\bfseries \mFid w/ $\bm{\erad = 1.0}$ (\texttt{eps1})} \hspace{.25\linewidth}
              \textsf{\bfseries \mFid w/ $\bm{\erad = 0.01}$ (\texttt{eps001})} \\ \vspace{1.3mm}
    \includegraphics[width=0.45\linewidth]{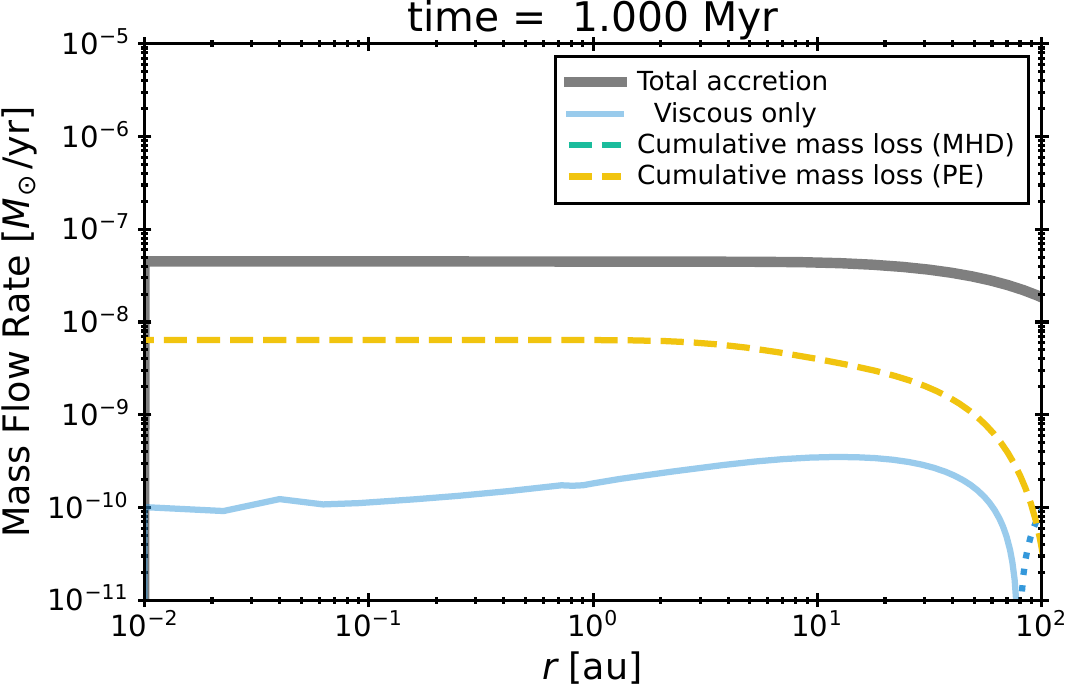}
     \includegraphics[width=0.45\linewidth, trim={0mm 0mm 0 0mm},clip]{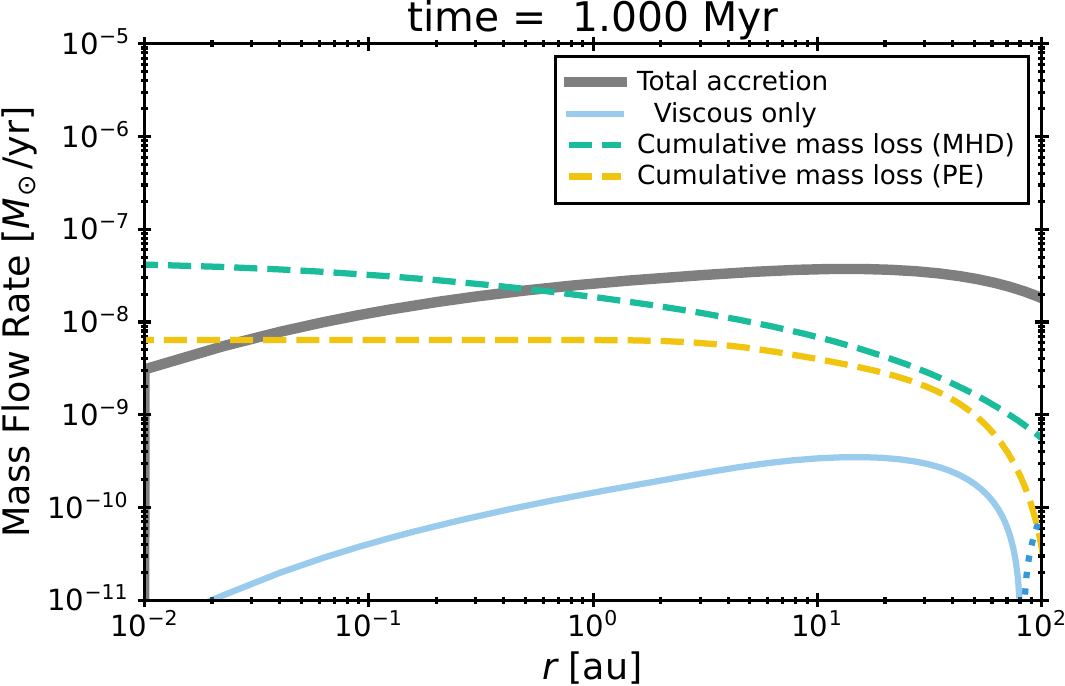}
    \caption{
    Distribution of the mass accretion and loss rates as in Fig. \ref{fig:Mdot_fid}, but for $\erad = 1$ (\texttt{eps1}; left) and $\erad = 0.01$ (\texttt{eps001}; right).
    \label{fig:Mdott_1myr_fheat}
    }
\end{figure*}

\begin{figure*}
    \centering
              \hspace{0.05\linewidth}
              \textsf{\bfseries \mFid w/ $\bm{\erad = 1.0}$ (\texttt{eps1})} \hspace{.25\linewidth}
              \textsf{\bfseries \mFid w/ $\bm{\erad = 0.01}$ (\texttt{eps001})} \\ \vspace{1.3mm}
    \includegraphics[width=0.48\linewidth]{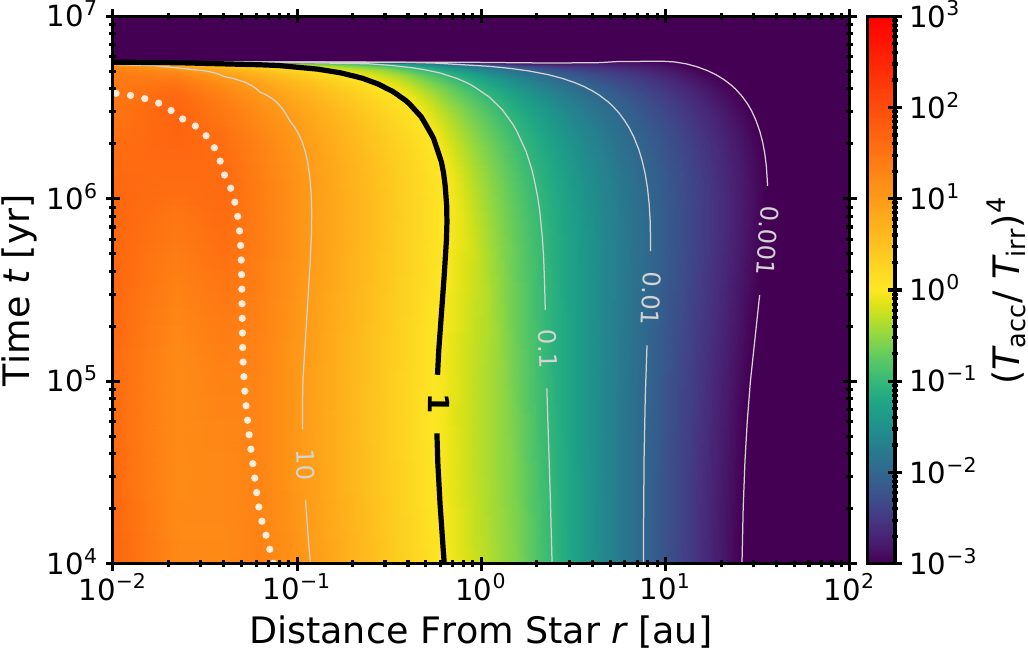}
        \includegraphics[width=0.48\linewidth]{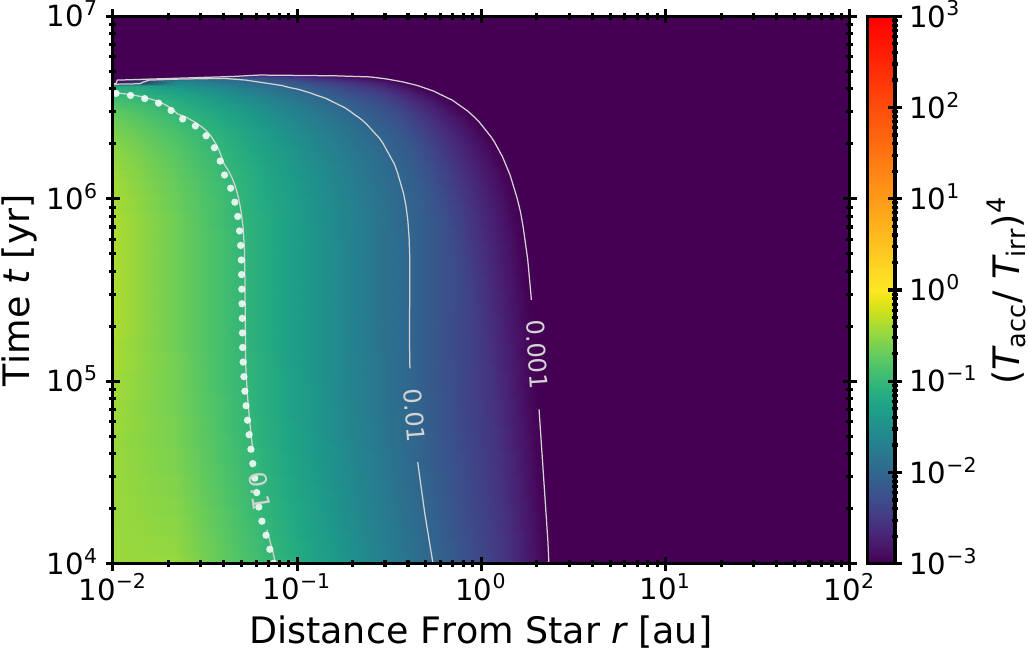}
    \caption{
    Space--time distribution of $(\Tacc/\Tirr)^4$ as in Fig. \ref{fig:Tacc_fid}, but for $\erad = 1$ (\texttt{eps1}; left) and $\erad = 0.01$ (\texttt{eps001}; right).
    \label{fig:Tmid_rt_fheat}
    }
\end{figure*}

Although our model depends on $\erad$, the representative value is uncertain.
This parameter represents the fraction of accretion energy consumed in the disk.
Thus, it would depend on the details of the energy dissipation process.
The local MHD simulations by \citetalias{Mori2021Evolution-of-th} showed that $\erad$ varies between 0.5 and $5\times10^{-4}$.
In particular, $\erad$ depends on the direction of the global magnetic field penetrating the disk.
The parameter $\erad$ not only controls the temperature distribution, but can also affect the mass-loss rate and disk evolution.
We varied $\erad$ to $\erad = 1$ and $\erad = 0.01$ and investigated their influence on the disk evolution.

Figure \ref{fig:Mdott_1myr_fheat} shows the evolution of the accretion and mass-loss rates.
When $\erad = 1$, all the accretion energy is dissipated in the disk; thus, no mass is lost due to the MHD disk wind (see Eq. (\ref{eq:cwe})).
Thus, the accretion rate is uniform within 10 au.
This case corresponds to the assumption by \citetalias{Mori2021Evolution-of-th}, where $\erad = 1$ and the mass loss is negligible.
For smaller $\erad$, the accretion rate decreases as in \mFid, but its distribution does not vary even though $\erad$ is smaller than that in \mFid.
This is because the mass-loss rate is not affected by the change in $\erad$ when $\cwe$ is larger than $\cwo$ in our model.

The space--time diagrams of the accretion-heated region are shown in Fig. \ref{fig:Tmid_rt_fheat}.
For the case of $\erad = 1$, the contribution of accretion heating to the midplane temperature is higher than in \mFid because the energy removed by the wind is used in heating.
Nevertheless, even in this extreme case, the accretion heating is weaker than that in the classical case (\mEff; see Fig. \ref{fig:Tmid_rt_fheat}).
For smaller $\erad$, the accretion heating becomes subdominant even within 0.01 au by further reducing the heating energy from \mFid.

\subsubsection{Influences of dust growth}
\label{ssec:dust-effect}

Next, we investigated the influence of the dust model on the disk structures.
The temperature structure in the MHD model depends on the distribution of the opacity and ionization fraction, which are determined by dust properties (e.g., size distribution).
As shown by \citet{Kondo2023The-Roles-of-Du}, the dust growth varies the ionization fraction and opacity, and it thereby affects the strength of the accretion heating.
Our model has two parameters for the dust properties: the dust abundance, which controls the ionization fraction, and the infrared opacity, which controls the accretion heating temperature (see Sect. \ref{sec:Tacc}).
Our fiducial model used the same dust model as in \citetalias{Mori2021Evolution-of-th}, where the abundance and size of dust grains are like those of interstellar dust.
We introduce another dust model that is based on the results with a maximum size of 100 $\mu$m, as iby \citet{Kondo2023The-Roles-of-Du}.
This case corresponds to the case in which accretion heating is maximized by dust growth.
To model their results, we chose a dust abundance of $10^{-3}$ by fixing the total surface area with their calculation and a reference opacity of $5 $ cm$^{2}$ g$^{-1}$.

\begin{figure}
    \centering
     \textsf{\bfseries  \mFid w/ dust growth effects (\mDG)}  \\ \vspace{1.3mm}
    \includegraphics[width=0.95\linewidth]{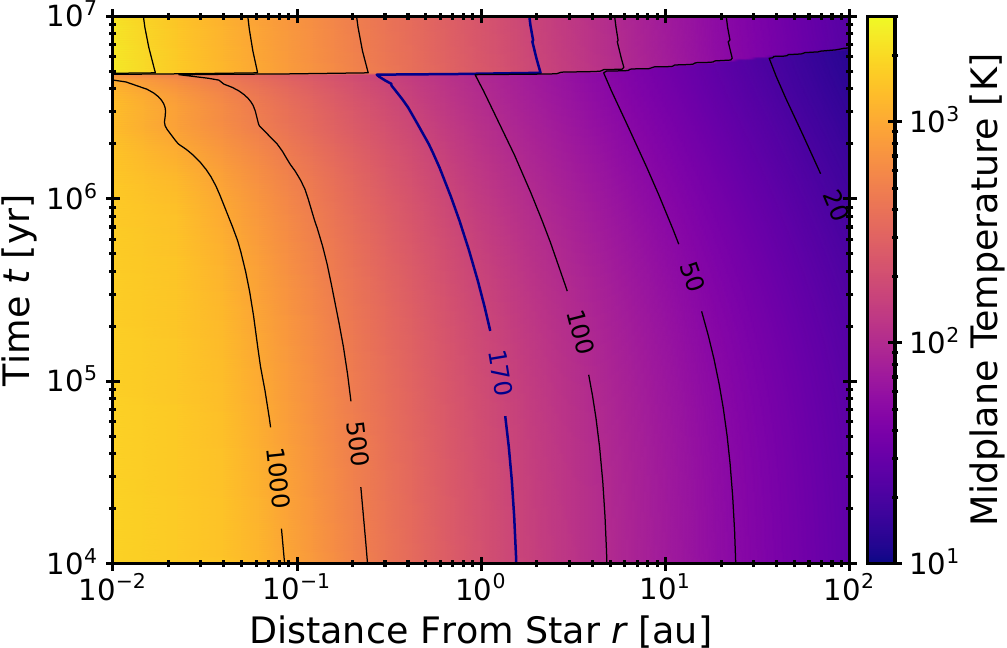} \\
       \includegraphics[width=0.97\linewidth]{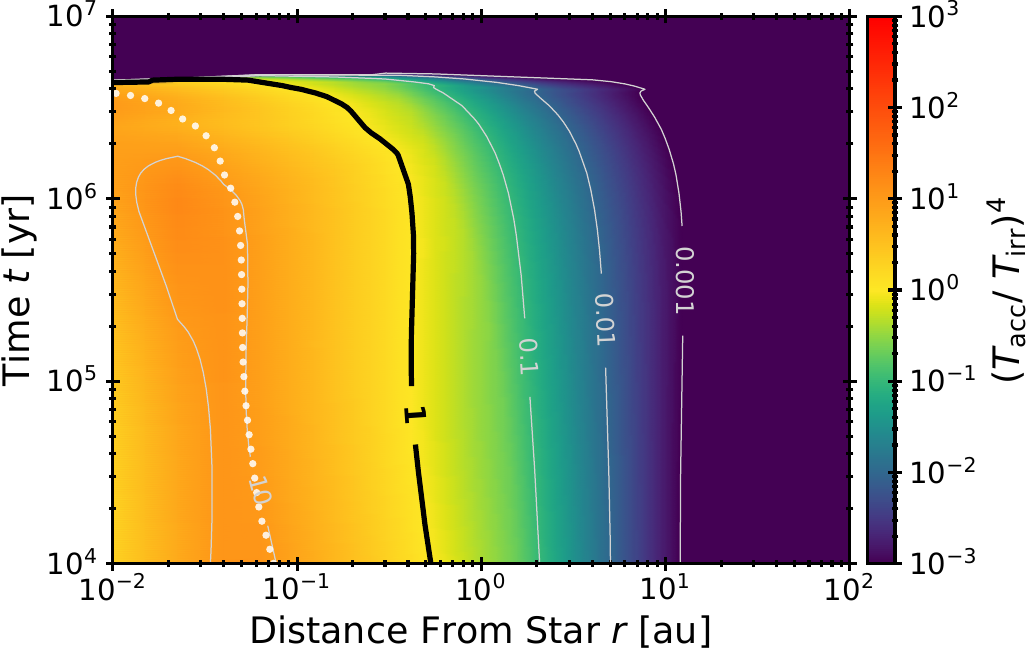}
    \caption{Space--time distribution of $\Tmid$ (top) and $(\Tacc/\Tirr)^4$ (bottom) for \mDG, which considers the effects of dust growth.
    \label{fig:Tmid_rt_dust}
    }
\end{figure}

Figure \ref{fig:Tmid_rt_dust} shows the space--time distribution of $\Tmid$ and $(\Tacc/\Tirr)^4$.
In this model, the accretion-heated region extends within $\sim$ 0.3 au, which is larger than that in \mFid.
However, the evolution of the snowline remains almost unchanged, although \citet{Kondo2023The-Roles-of-Du} showed that snowline migration could be delayed by dust growth.
This is because in this model, $\erad$ is set to a lower value of 0.1; thus, accretion heating is sufficiently weakened (see Sect. \ref{ssec:dis-sl}).
This suggests that although accretion heating may be enhanced by dust growth, this effect depends on the value of $\erad$.

\subsubsection{Effects of magnetic flux transport}
\label{sec:apz-effect}

\begin{figure}
    \centering
     \textsf{\bfseries \mFid w/ $\bm{\Sigma}$-dep $\bm{\apz}$  (\mApz)}  \\ \vspace{1.3mm}
\hspace{-0.12\linewidth}
    \includegraphics[width=0.86\linewidth]{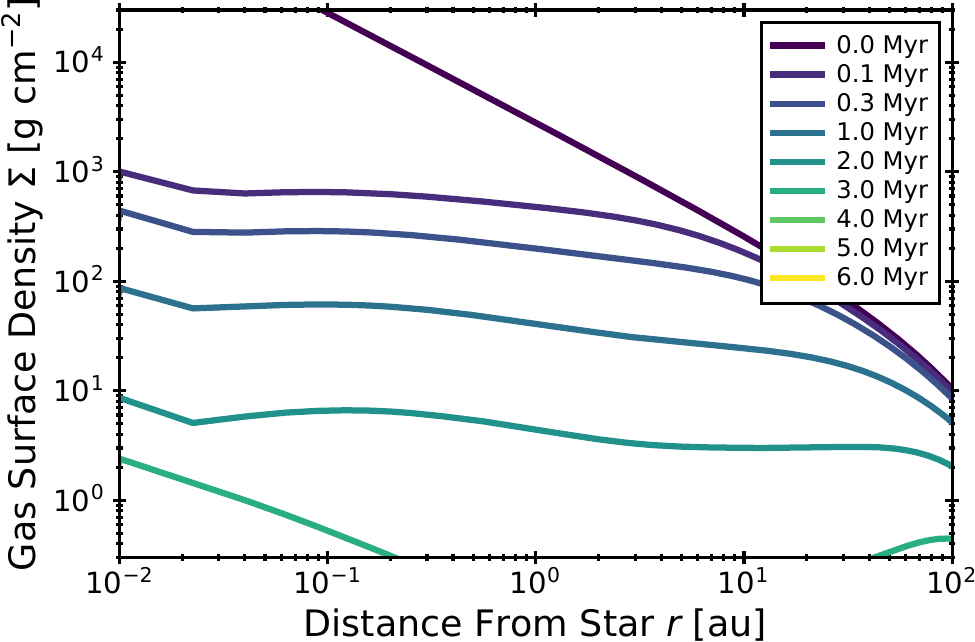}
          \includegraphics[width=0.95\linewidth]{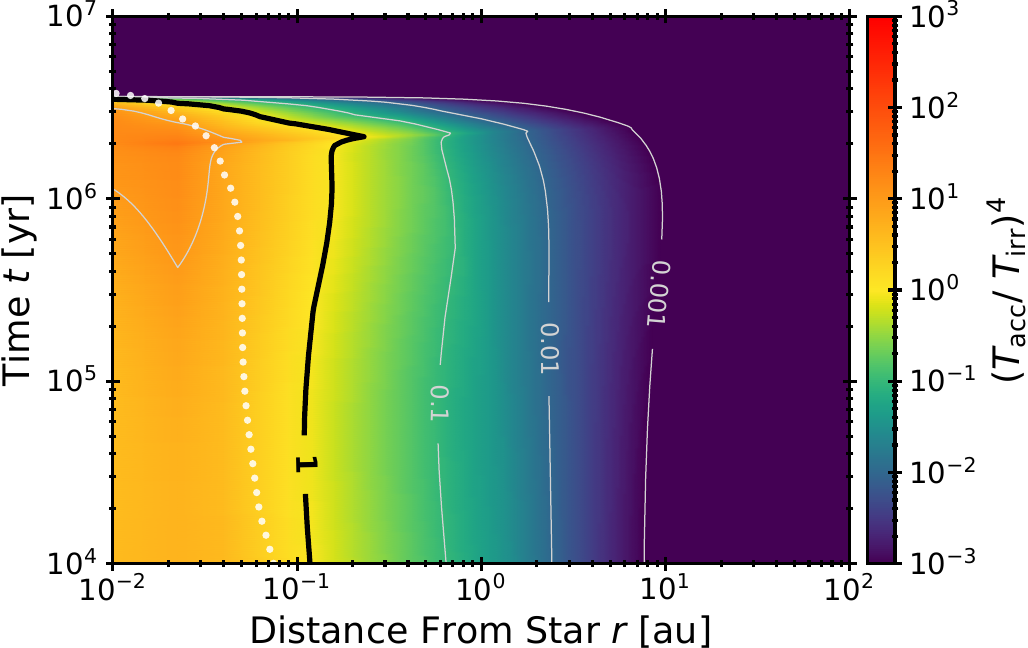}
    \caption{
    Left: Evolution of the surface density profile for \mApz, where $\apz$ depends on $\Sigma$.
    Right: Space--time distribution of $(\Tacc/\Tirr)^4$ as in Fig. \ref{fig:Tacc_fid}, but for \mApz.
    \label{fig:mdoel-Aa}
    }
\end{figure}

We examined the effect of magnetic flux transport on the disk structure by considering the case where the magnetic stress depends on the gas surface density $\Sigma$.
The time evolution and profile of $\apz$ are sensitive to how magnetic flux is transported during disk evolution.
If the magnetic flux is transported along with the gas accretion, the relative importance of the magnetic field can be maintained to some extent \citep{Lubow1994Magnetic-field-,Okuzumi2014Radial-Transpor}.
Despite uncertainties in the flux transport, we assumed $\apz$ to be constant in the fiducial model (see Sect. \ref{ssec:setting}).
When the magnetic flux is not transported, the magnetic flux relative to the surface density is increased by the decrease in the surface density \citep[e.g.,][]{Bai2013bWind-driven-Acc}.
This enhances the magnetic stress.
This $\Sigma$-dependent $\apz$ model, based on the local MHD simulations \citep{Bai2013bWind-driven-Acc}, was described by \citet{Suzuki2016Evolution-of-pr}.
We adopted the $\apz$ model,
\begin{equation}
    \apz =  \min \pr{10^{-3} \pf{\Sigma}{\Sigma_{\rm ini}}^{-0.66} , 1 }
\end{equation}
where $ \apz$ increases as $\Sigma$ decreases.

Figure \ref{fig:mdoel-Aa} shows the evolution of $\Sigma$.
In the first 0.1 Myr, the surface density rapidly decreases because the disk accretion due to $\apz$ strengthens over time.
When the accretion rate is higher than the photoevaporative mass-loss rate, gas accretion from the outer region fills the gap and maintains the surface density to some extent \citep{Kunitomo2020Dispersal-of-pr}.
This explains why the surface density evolution is more gradual in the late stage than in \mFid.

Figure \ref{fig:mdoel-Aa} also shows the space--time distribution of $(\Tacc/\Tirr)^4$.
The accretion-heated region lies within $\sim 0.1$ au.
Thus, even with the change in the $\apz$ evolution, our conclusion that accretion heating is not the dominant heating source in disk evolution remains robust.

The accretion-heated region gradually shifts outward and peaks around $t =$ 2 Myr.
This occurs because the heating altitude approaches the midplane as $\Sigma$ decreases because of gas accretion.
In previous cases, this behavior was not observed because $\Sigma$ rapidly drops because of photoevaporation.

\subsection{Influences on planet formation}
\label{ssec:infl-plf}

We demonstrate the influence of the disk temperature model on planet formation processes by comparing typical cases of MHD heating (\mFid) and conventional heating (\mEff).
We calculated the mass-orbital track for protoplanets using time-dependent disk structures obtained in calculations.
We considered protoplanet growth through pebble accretion and gas accretion and migration through gravitational interaction (i.e., migration types I and II) with the disk gas, as described in detail below.
Furthermore, considering the evolution of the water fraction, we addressed the origin of the observed dichotomy \citep[e.g.,][]{Parc2024From-super-Eart} in the water fraction of close-in planets in the mass range of 1--10 $M_\oplus$ (i.e., super-Earths and sub-Neptunes).

\begin{figure*}
    \centering \figtitlen{0.2}{0.12}
    \includegraphics[height=0.35\linewidth, trim={0 0mm 0 0mm},clip]{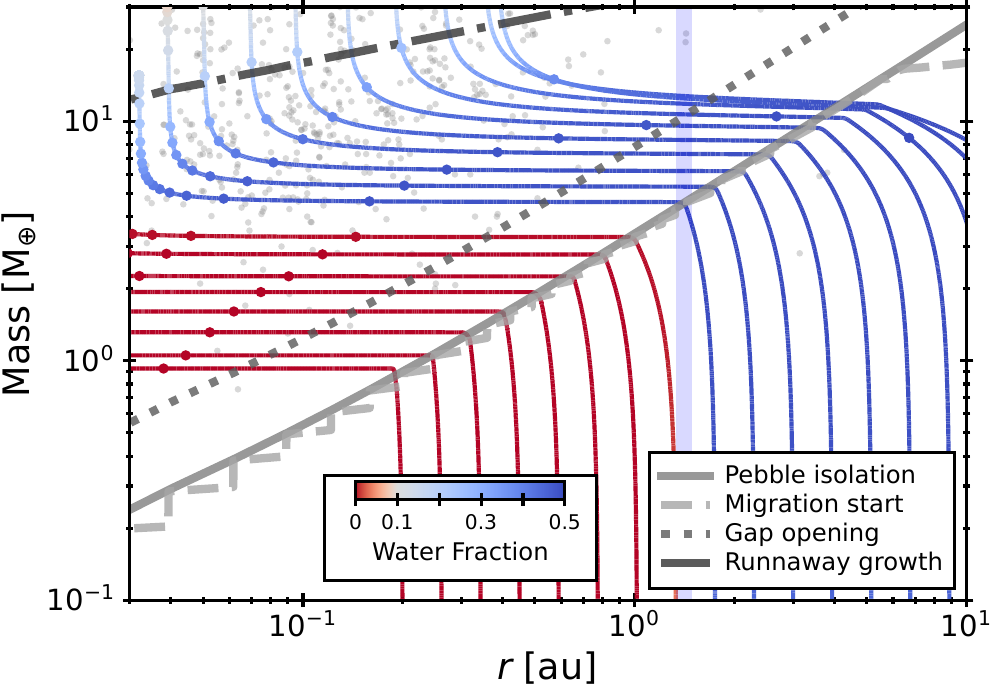}
    \includegraphics[height=0.35\linewidth, trim={20mm 0mm 0 0mm},clip]{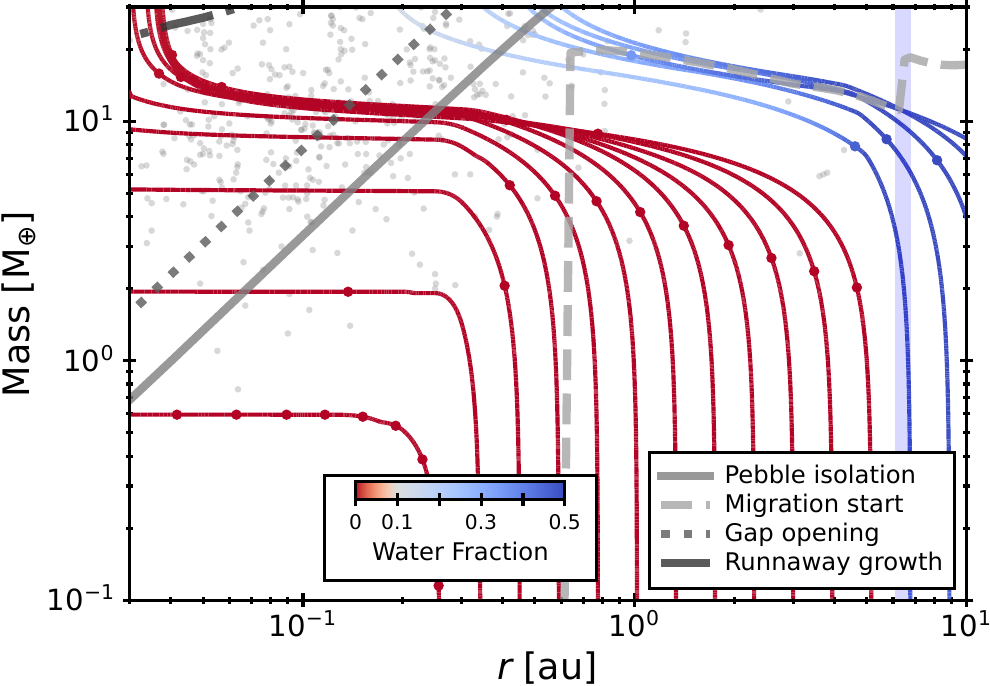}
   \caption{
        Mass--orbital tracks of protoplanets starting at $\Mp = $0.01 $M_\oplus$ and $r = [0.2{:}20]$ au, with early-onset growth ($t_0 = 0.01$ Myr) and $v_{\rm frag} = 10$ m s$^{-1}$, under MHD heating (left) and classical heating (right). The total pebble mass drifting through the snowline at $t \ge t_0$ is 400 $M_\oplus$.
   The characteristic masses in planetary evolution at $t = 0.1$ Myr are also shown as gray lines.
    Planets grow through pebble accretion until they reach the isolation mass ($\Miso$; solid line).
    When the growth timescale exceeds the migration timescale, migration becomes the dominant process ($\Mmig$; dashed line).
    Gaps start to open when $\Mp$ reaches $2.3 \Miso$ ($\Mgap$; dotted line).
    Planets grow in a runaway manner when the growth timescale again becomes shorter than the migration timescale ($\Mrg$; dash-dotted line).
    The points on each mass-orbital track show the elapsed time for every 0.5 Myr.
    The gray circles show the distribution of confirmed exoplanets orbiting stars with masses in the range of $[0.8, 1.2] M_{\odot}$ \citep{NASA-Exoplanet-Science-Institute2020Planetary-Syste}.
    The vertical blue lines show the location of the snowline at $t = 0.1$ Myr.}
    \label{fig:masses}
\end{figure*}

\subsubsection{Model: Growth rate and migration rate}
\label{ssec:model-growth}

For pebble accretion, we followed the model described by \citet{Liu2019aSuper-Earth-mas} and \citet{Ormel2018Catching-drifti}.
The pebble accretion rate $\dot{M}_{\rm PA}$ is given as
\begin{align}
    \label{eq:Mdot_peb}
\dot{M}_{\rm PA} &=
        \begin{dcases}
             \pr{ \varepsilon_{\rm 2D}^{-2} + \varepsilon_{\rm 3D}^{-2}}^{-0.5} ~ \dot{M}_{\rm peb} & (\Mp<\Miso) \\
            0 & (\Mp \ge \Miso)
        \end{dcases} , \\
\varepsilon_{\rm 2D} &= \frac{0.32  }{ |u_{\rm dust}|/(2v_{\rm K} {\rm St})  }  \sqrt{ \frac{\qp \Delta v}{ {\rm St} ~ v_{\rm K} }},\\
\varepsilon_{\rm  3D} &= \frac{0.39 \qp  \sqrt{1 + {\rm St}/\alphat } }{h  |u_{\rm dust}| /(2v_{\rm K} {\rm St}) }  ,\\
\Delta v &=  \pr{ 1 + 5.7 \qp \, {\rm St}\, \eta^{-3}  }^{-1} \eta v_{\rm  K} + 0.52 \pr{\qp {\rm St}}^{1/3} v_{\rm  K} ,
\end{align}
where $M_{\rm p}$ is the protoplanet mass,
$\Miso$ is the pebble isolation mass,
$\dot{M}_{\rm peb}$ is the pebble mass flux in the disk,
$\qp = M_{\rm p}/M_\star$,
$h = H/r $,
$u_{\rm dust } = - 2 \eta v_{\rm K} \St + v_{\rm gas}$, with $v_{\rm gas}$ being the radial gas velocity,
the pebble Stokes number $\rm St \lesssim 1$, and we assumed that turbulence with the strength $\alphat$ controls the dust diffusion.
We took the gas accretion velocity in the drift speed of dust relative to protoplanets into account as in \citet{Liu2019aSuper-Earth-mas}.
The radial gas velocity $v_{\rm gas}$ was assumed to $- \dot{M}_{\rm acc}/(2 \pi r \Sigma)$, although it can be reduced for an efficient surface accretion \citep{Okuzumi2025Surface-accreti}.

To provide $\St$, we took the approach with the parameterization of the pebble flux $\dot{M}_{\rm peb}$, where the pebble flux parameter $\xi \equiv \dot{M}_{\rm peb}/\dot{M}_{\rm acc}$ as in \citet{Ida2016The-radial-depe}, and
\begin{equation}
\xi =
        \begin{dcases}
            \xi_0 & (T < 170\, \K) \\
            \xi_0/2 & (1500\, \K > T > 170\, \K) \\
            0 & (T > 1500\, \K)
        \end{dcases}
        ,
\end{equation}
where the ice and silicate sublimation were modeled.
The reference pebble flux parameter $\xi_0$ is given from fitting to the result of a dust transfer simulation in Appendix \ref{app:pebble-twopop} \citep{Birnstiel2012A-simple-model-},
\begin{equation}
\label{eq:fit-xi-eq}
        \xi_0 =  0.15 \exp\left[ - \pr{ t / 0.05\,{\rm Myr} }^{0.52} \right] .
\end{equation}
The pebble distributions were assumed to be in a steady state without growth calculations.
The Stokes number is given by the size limits of dust growth, radial drift, turbulence-induced fragmentation, and drift-induced fragmentation based on \citet[][]{Drazkowska2021How-dust-fragme}, while we slightly updated the drift limit to be applicable for small $\St$ (see Appendix \ref{app:pebbleSt}),
\begin{align}
\mathrm{St} &= \min\pr{ \mathrm{St}_{\mathrm{drift}},\, \mathrm{St}_{\mathrm{frag}},\, \mathrm{St}_{\mathrm{df}},\, \mathrm{St}_{\mathrm{growth}} } \label{eq:St} \\
\mathrm{St}_{\mathrm{drift}} &= \frac{v_{\mathrm{gas}}}{4 \eta v_{\rm K}}
        + \sqrt{  \pf{v_{\mathrm{gas}}}{4 \eta v_{\rm K}}^2
        + 0.55\, \frac{\dot{M}_{\mathrm{peb}}} {8\pi  \Sigma_g r v_{\rm K} \eta | \eta |} } , \label{eq:St-drift} \\
\mathrm{St}_{\mathrm{frag}} &= 0.37 \, v_{\rm frag}^2/ \pr{3  \alpha_t \cs^2}, \\
\mathrm{St}_{\mathrm{df}}   &= v_{\rm frag} / (\lvert \eta \rvert v_{\rm K}) , \\
\mathrm{St}_{\mathrm{growth}} &= \frac{\pi \rho_{\rm s} a_0}{2 \Sigma} \exp \pr{ t \Omega \epsilon_0 } ,
\end{align}
where
 the monomer size $a_0$ was set to $0.1\,\mu {\rm m}$,
 the grain internal density $\rho_{\rm s}$ was set to $1.3\, {\rm g\, cm^{-3}}$,
 the initial dust-to-gas mass ratio $\epsilon_0$ was set to $0.01$,
 and we adopted the fitting parameters of a dust coagulation simulation \citep{Birnstiel2012A-simple-model-}.
The fragmentation velocity $v_{\rm frag}$ was set to 10 m s$^{-1}$ for the fiducial case, while the case with $v_{\rm frag} = 1$ m s$^{-1}$ is shown in Appendix \ref{app:tracks}.
We also confirmed that the pebble profiles precisely matched the result of a two-population dust transport simulation \citep{Birnstiel2012A-simple-model-} under a given pebble flux (Appendix \ref{app:pebbleSt}).

Pebble accretion halts when a protoplanet reaches the pebble isolation mass $\Miso$ \citep{Bitsch2018Pebble-isolatio,Pichierri2024A-Recipe-for-Ec},
\begin{equation}\label{eq:Miso}
        \Miso = 25 M_\oplus \pf{h}{0.05}^3 \!  \left[  0.34 \pf{ {-3} }{ \log(\alpha)  }^4 \! \! + 0.66 \right]
        \pr{\frac{3.5 +\chi}{6} }.
\end{equation}
The original formula was based on viscous disks.
We simply assumed $\alpha$ to be $\alphat=10^{-4}$ because the proper treatment for $\alpha$ in laminar magnetized disks is unclear.
We note that the pebble-isolation mass is proportional to $H^3$ \citep{Bitsch2018Pebble-isolatio}.
As shown by \citet{Bitsch2019Inner-rocky-sup}, colder disks show lower $\Miso$, whereas hotter disks show higher $\Miso$, which impacts planetary formation processes.

For the protoplanet growth rate via gas accretion, we followed \citet{Johansen2019How-planetary-g} \citep{Ikoma2000Formation-of-Gi,Tanigawa2016Final-Masses-of},
\begin{align}
\label{eq:Mdot_gas}
        \dot{M}_{\rm gas} &= \min\pr{ \dot{M}_{\rm KH}, \dot{M}_{\rm disk}, \dot{M}_{\rm acc} } , \\
        \dot{M}_{\rm KH} &= 10^{-5} M_\oplus {\rm yr}^{-1} \pf{\Mp}{10 M_\oplus}^4 \pf{\kappa_{\rm pl}}{1\, {\rm cm}^2 \, {\rm g}^{-1}}^{-1} , \\
        \dot{M}_{\rm disk} &= 0.29 h^{-2} \Mp^{4/3} \Sigma r^2 \Omega \left[1 + \pr{\Mp / M_{\rm gap} }^2 \right]^{-1} ,
\end{align}
where
the opacity in the envelope of the protoplanet, $\kappa_{\rm pl}$, was set to 1 ${\rm cm}^2 {\rm g}^{-1}$,
and $M_{\rm gap}$ is the planetary mass at which the gap opens in the disk.
By increasing the protoplanet mass, the protoplanet begins to scatter the disk gas, and the gap opens in the disk.
As discussed by \citet{Johansen2019How-planetary-g}, the gap in the disk should be opened after the protoplanet arrives at the pebble isolation mass.
We also simply adopted their prescription for the gap opening mass, that is, $M_{\rm gap} = 2.3 M_{\rm iso}$.
Although this gap depth function differs slightly from those in studies that modeled a gap opening \citep{Kanagawa2018Radial-Migratio,Pichierri2023A-recipe-for-or, Pichierri2024A-Recipe-for-Ec} and may influence planetary evolution, we confirmed that it does not qualitatively affect our conclusions.

To obtain the orbital migration rate of protoplanets, we calculated the formulae of planetary torque $\Gamma$ given by \citep{Paardekooper2011A-torque-formul}, which consider viscosity and thermal diffusion to retain the corotation torque. For a given $\Gamma$,  the orbital migration rate is calculated by
\begin{equation}
    \dot{r}_{\rm mig} = 2  \frac{\Gamma}{\Mp r \Omega} \left[1 + \pr{\Mp / M_{\rm gap} }^2 \right]^{-1} ,
\end{equation}
where the factor $\left[1 + \pr{\Mp / M_{\rm gap} }^2 \right]^{-1}$ is introduced to suppress the migration rate when the gap is opened, as in \citet{Johansen2019How-planetary-g}.

\subsubsection{Mass-orbital track}

We placed protoplanets with $\Mp =$ 0.01 $M_\oplus$ at $t_0 = 0.01$ Myr (the onset time of protoplanet growth) in the disk evolution models with MHD heating (left) and classical heating (right).
The obtained mass-orbital tracks are shown as lines in Fig. \ref{fig:masses}.
We also show the water fraction of the protoplanets.
We assumed that the water fraction of pebbles is 50\% outside the snowline and zero inside the snowline.
We also assumed that the water fraction of the disk gas is zero outside the snowline and $0.005$ inside the snowline \citep{Lodders2003Solar-System-Ab}.

To better understand the growth and migration of protoplanets, we also show characteristic masses based on the disk structures at $t = t_0$.
The first mass is the pebble isolation mass $\Miso$ (solid line) as described in Eq. (\ref{eq:Miso}).
The second mass is the migration initiation mass $\Mmig$ (dashed line), which is the mass at which the migration timescale $(r/\dot{r}_{\rm mig})$ becomes shorter than the growth timescale $(\Mp/\dot{M}_{\rm peb})$.
The third mass is the gap-opening mass $\Mgap$ ($=2.3\Miso$; dotted line), which is the mass at which the gap opens in the disk.
The last mass is the runaway growth mass $\Mrg$ (dash-dotted line), which is the mass at which the growth timescale ($=\Mp/\dot{M}_{\rm gas}$) again becomes shorter than the migration timescale.

We begin with the case of the classical heating model (\mEff).
In this model, the high disk temperature leads to a high pebble isolation mass.
In this parameter set for young PPDs, the protoplanets efficiently grow via pebble accretion due to the high pebble mass flux.
While the migration initiation mass is higher than $10\,M_\oplus$ at $t = t_0$, it decreases as the pebble flux decreases.
Thus, the protoplanets begin to migrate inward around $M \sim 4$--10 $M_\oplus$ before they reach the pebble isolation mass.
The inner protoplanets do not grow until the disk temperature decreased enough ($T < 1500$ K).

On the other hand, the colder disk model (\mFid) shows a distinct evolutionary path.
The reduced aspect ratio $h$ lowers the pebble isolation mass and increases the pebble accretion efficiency due to the inverse dependence on $\eta $ ($\propto h^2$).
As a result, protoplanets rapidly reach the isolation mass and begin to migrate inward as the pebble accretion ceases.
Although the gap opening slows this migration (i.e., type II migration), lower-mass protoplanets ($M_{\rm p} \lesssim 4\,M_\oplus$) still migrate inward on short timescales.
To match the observed exoplanet distribution, additional disk structures are required to halt or slow this migration.
Meanwhile, more massive protoplanets ($M_{\rm p} \sim\,10 M_\oplus$) undergo runaway gas accretion and form close-in gas giants.

The differences in the disk models affect the water content of protoplanets.
In model \mEff, a high $\Miso$ and distant snowline lead to fully rocky super-Earth or sub-Neptune protoplanets.
In contrast, model \mFid, with a lower $\Miso$ and a closer-in snowline, shows a transition from rocky to water-rich compositions around $\Mp \sim 4\,M_\oplus$.
This agrees with the observation that indicated a compositional transition near $4.2\,M_\oplus$ for FG-type stars \citep{Parc2024From-super-Eart}.
Although other parameter sets in Appendix \ref{app:tracks} show varying transition masses, the presence of the transition around a few $M_\oplus$ is less sensitive to disk parameters than in \mEff.
Even when the transition occurs at lower masses, collisional merging, which was neglected here, may shift the transition mass higher.

Even in viscous disks, the observed water transition may occur if the planetary growth timescale is comparable to the inward migration timescale at $\Mp \sim$ a few $M_\oplus$, as in previous studies \citep[e.g.,][]{Venturini2020Most-super-Eart, Izidoro2021Formation-of-pl, Izidoro2022The-Exoplanet-R}.
For instance, as shown in Appendix~\ref{app:tracks} (panel (d) of Fig.~\ref{fig:app-tracks}), a case with a later onset of protoplanet growth ($t_0 = 0.2 \mathrm{Myr}$) shows a transition in the water fraction around $\Mp \sim 4 M_\oplus$ even in \mEff.
Furthermore, as in \citet{Venturini2020Most-super-Eart}, lower disk masses can lead to a modest reduction in pebble accretion efficiency.
Additional migration torques, such as angular momentum changes due to pebble accretion \citep{Lambrechts2019Formation-of-pl, Izidoro2021Formation-of-pl} and thermal torques \citep{Guilera2019Thermal-torque-,Venturini2020Most-super-Eart}, alter the migration timescale and influence the conditions under which the water fraction transition at a few $M_\oplus$ occurs.

Nevertheless, we emphasize that
when the pebble isolation mass is as low as in our models,
efficient pebble accretion alone can account for the coexistence of rocky and volatile-rich planets for close-in planets.
The early onset of protoplanet growth is also consistent with observed disk mass statistics, which suggest that planet formation likely begins early in the disk evolution \citep[e.g.,][]{Andrews2020Observations-of}.
Furthermore, we explore in Appendix \ref{app:tracks} cases in which $v_{\rm frag}$ and $t_0$ were varied. The transition in the water fraction then occurs at about a few Earth masses.
Taking into account collisional merging may further help us to reproduce the observed water-fraction transition.
In contrast, for viscous disks with a higher pebble isolation mass, a reproduction of the water-fraction transition requires a fine-tuning of model parameters.

The results also provide implications for the formation of Earth.
If a protoplanet with $\Mp \sim $ 0.01 $M_\oplus$ existed at 1 au, it would quickly reach the pebble isolation mass of higher than $1\,M_\oplus$ even when the pebble isolation mass is low (\mFid).
To avoid this, the pebble flux should have been suppressed well during the formation of Earth.
This might be due to the halting of pebble drift by the Jupiter-created gap or the depletion of dust in the disk.

Our calculation neglected planetary growth through collisions with protoplanets or planetesimals.
If collisional growth is significant, protoplanets may acquire additional mass, which might lead to the formation of larger planetary bodies.
In addition, gap formation by protoplanets in the outer region can influence the pebble flux delivered to the inner disk.
Furthermore, recycling of vapor around protoplanets \citep{Johansen2021A-pebble-accret,Wang2023Atmospheric-rec,Muller2024A-formation-pat} and heating by the $^{26}$Al radionuclide in planetesimal-sized bodies \citep{Lichtenberg2019A-water-budget-,Lichtenberg2021Bifurcation-of-} can suppress the planetary water content and might facilitate the formation of rocky planets in the later phases of disk evolution.
Detailed studies of planet formation within more comprehensive models are therefore required.
Nonetheless, our study highlights the role of pebble accretion alone in shaping the water-fraction distribution across planetary masses.

\section{Discussion} \label{sec:discussion}

\subsection{Influences of energetics}

\begin{figure}
    \centering
        \hspace{-0.04\linewidth}
    \includegraphics[width=0.85\linewidth]{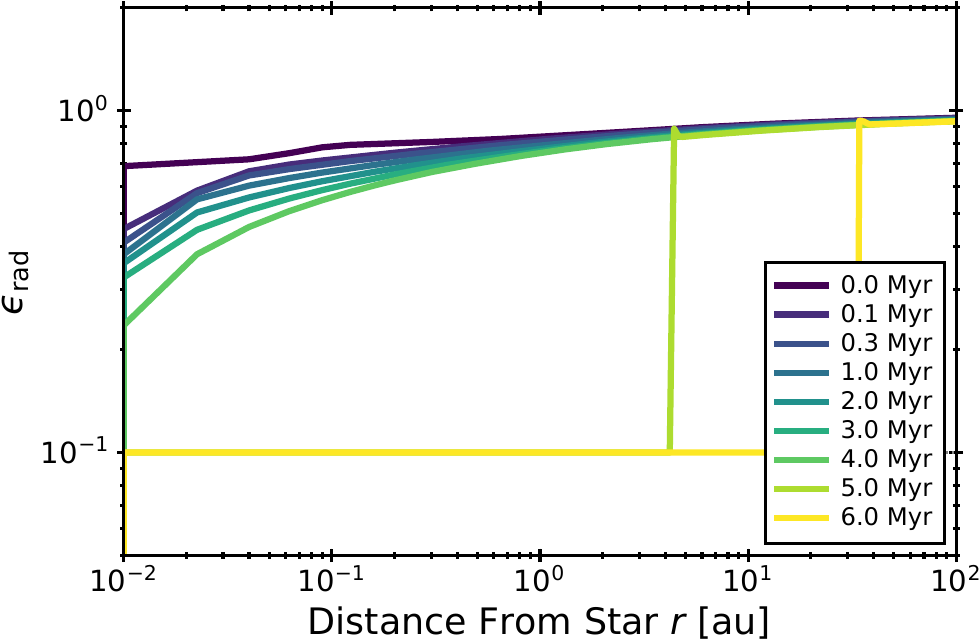} \\
    \hspace{0.04\linewidth}
       \includegraphics[width=0.95\linewidth]{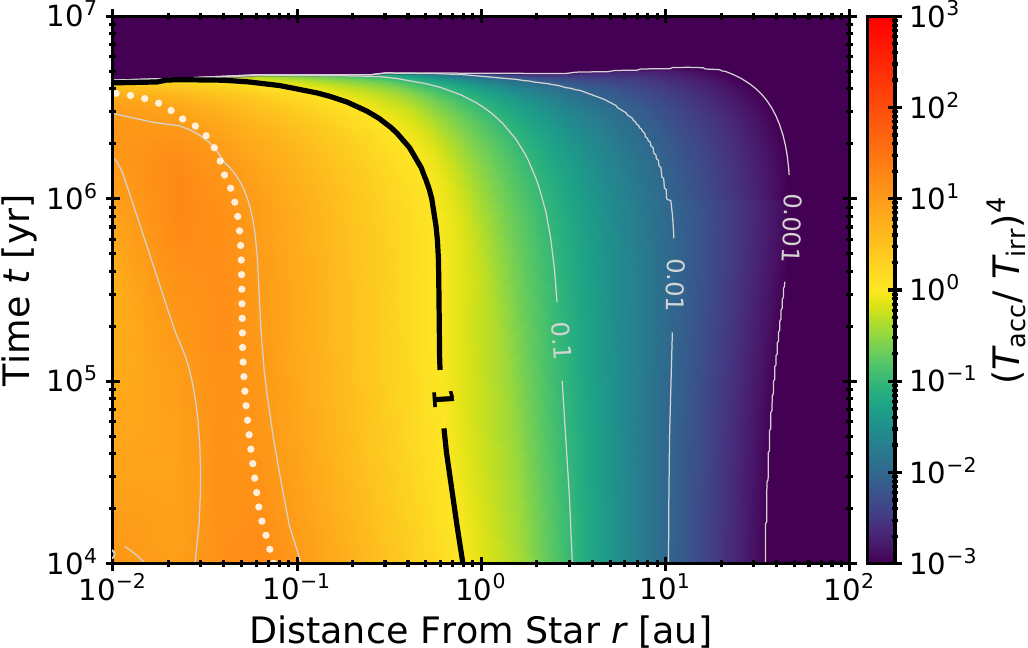}
    \caption{Space--time distribution of $\erad$ (top) and $(\Tacc/\Tirr)^4$ (bottom) as in \mFid, but modified to meet the energy balance.
    \label{fig:eradTacc_rt_EB}
    }
\end{figure}

We used the parameter $\cwo$ to determine the mass-loss rate of MHD winds.
Simultaneously, $\erad$ determines the energy that is used for the accretion heating.
The generated and lost energies are not the same.
When $\cwo$ is lower than $\cwe$, it implies some energy in the MHD wind.
However, these parameters were obtained from different local simulations that cannot capture the wind energy in the outer limit correctly.
Therefore, the actual balance between the wind energy and the heating energy is still highly uncertain.
As a possibility, while the wind energy (i.e., $E_{\rm w}$) is maintained low or zero, the rest of the energy may be used for the heating (see Eq. (\ref{eq:suzuki-enebalance})).

We demonstrate how much the unused energy affects the disk temperature.
At the radius where $\cwo > \cwe$, whereas the mass-loss rate is determined by $\cwo$ as in the fiducial cases (Eq. (\ref{eq:sigdotdw})), $\erad$ is increased so that the energy balance with the zero wind energy meets
\begin{equation}
         \Frad = \Grp+\Gpz - \SigdotDW r^2\Omega^2/2.
\end{equation}
We applied this approach to \mFid.
 Figure \ref{fig:eradTacc_rt_EB} shows the $\erad$ profile and the space--time distribution of $(\Tacc/\Tirr)^4$.
 $\erad$ deviates from 0.1 and $\sim$ 1 in the outer region at $t < $ of a few million years.
Since $\erad$ is 0.1 in \mFid, this suggests that most of the accretion energy is neglected in the outer region.
However, as in Sect. \ref{ssec:fheat-effect}, even with $\erad = 1$, the efficient cooling of the surface heating results in a cooler disk temperature than in \mEff.

The values to which $\cw$ and $\erad$ should be set are still unclear, although MHD simulations have measured the mass-loss rate \citep[e.g.][]{Suzuki2009Disk-Winds-Driv,Lesur2021Systematic-desc}.
The value of $\cw$ varies in different simulations; in particular, it is strongly dependent on the strength of the magnetic field \citep{Lesur2021Systematic-desc}.
In addition, $\cw$ obtained in simulations might be too high, which would result in inconsistencies with observations, such as the disk lifetime.
The evolution of the magnetic field must be revealed to understand the disk wind strength ($\apz$ as well; Sect. \ref{sec:apz-effect}).
Further research is necessary to determine the appropriate combination of $\cw$ and $\erad$ that is consistent with the energy balance.

\subsection{Comparison with observational $\dot{M}$--$t$ relations}
\label{ssec:disc-compobs}

\begin{figure}
    \centering
    \includegraphics[width=0.99\linewidth]{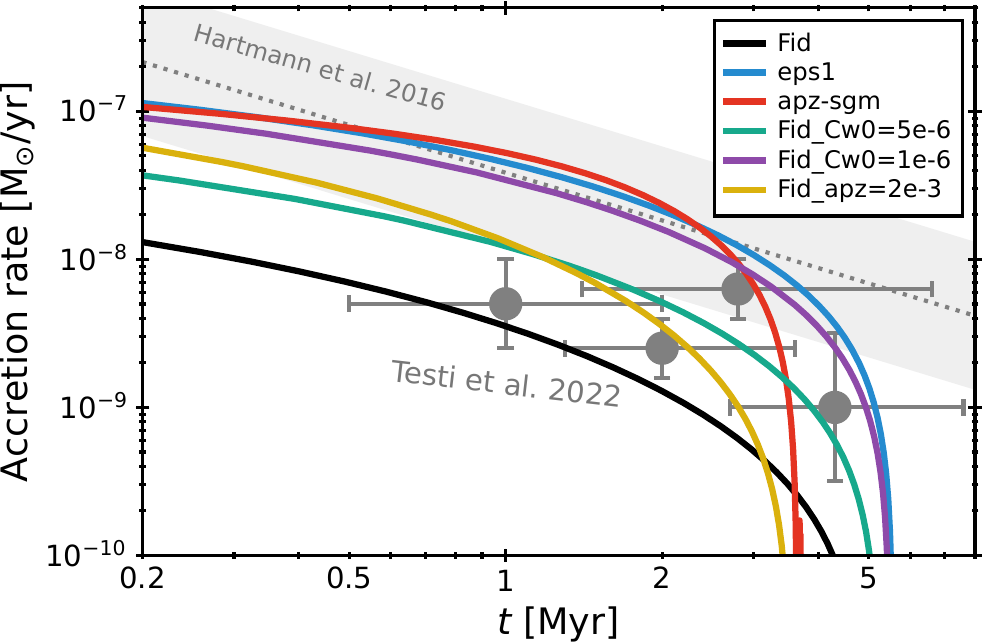}
    \caption{Evolution of the accretion rate in our simulations (solid lines) and the observational relation between the accretion rate and stellar age.
   The simulated accretion rates are measured at the inner boundary of the simulation domain.
    The observational $\dot{M}$--$t$ relations for $1\,M_\sun$ stars derived by \citealt[][]{Hartmann2016Accretion-onto-} (dotted gray line and shade) and \citealt[][]{Testi2022The-protoplanet} (dark gray points with bars) are shown.
    The horizontal bars show the 25th to 75th percentile range of the stellar ages in each star-forming region.
    \label{fig:Mdot-t_obs}
    }
\end{figure}

We discuss the consistency of our accretion rate evolutions with the observed relation between the stellar accretion rate and age \citep[e.g.,][]{Hartmann1998Accretion-and-t,Hartmann2016Accretion-onto-,Sicilia-Aguilar2010Accretion-in-Ev,Testi2022The-protoplanet}.
\citet{Hartmann2016Accretion-onto-} presented the empirical $\dot{M}$--$t$ relation\footnote{We note that although \citet{Hartmann2016Accretion-onto-} derived stellar ages with the original version of stellar evolutionary models by \citet{Feiden2016Magnetic-inhibi}, we shifted the time in our stellar luminosity evolution (see Sect. \ref{ssec:Tirr}).} for $M \sim 1\,\Msun$ stars,
$       \dot{M} = 4 \times 10^{-8\pm0.5}\,\smpy ~\pr{t/1\,{\rm Myr}}^{-1.07} , $
where the mass dependence was taken into account.
\citet{Testi2022The-protoplanet} also analyzed the accretion rate for different star-forming regions and derived the $\dot{M}$--$t$ relation.
Both relations are shown in Fig. \ref{fig:Mdot-t_obs}.
There may be a bias where stars with higher $\dot{M}$ are easily observed.
The observational relation still has a scatter on the accretion rate. This may depend on the observations and analyses.

The simulated $\dot{M}(r_{\rm in})$ evolutions are expected to intersect the observed relation at some point if the simulation reproduces typical disk evolutions.
Because the simulation only represents a single case, the evolution of its accretion rate does not need to follow observations fully.
Figure \ref{fig:Mdot-t_obs} also shows several $\dot{M}$ evolutions from our simulations.
The stellar accretion rate in \mFid deviates from Hartmann’s relation.
The reduced accretion rate in \mFid is caused by the mass loss by disk winds (see Fig. \ref{fig:Mdot_fid}).
However, the accretion rate evolution in \mFid is consistent with the mean values of two star-forming regions in \citet{Testi2022The-protoplanet}.
Thus, \mFid can explain stars with lower accretion rates.

Even in our simulations, higher accretion rates can be obtained by adjusting the model parameters of the disk within their uncertainties.
When $\erad$ is set to unity (see the model \texttt{eps1}), the accretion rate exceeds that of \mFid because there is no wind mass loss.
The evolutionary model with a $\Sigma$-dependent $\apz$ also maintains the higher accretion rate (see the model \texttt{apz-sgm}).
In both cases, the accretion rates are higher than in \mFid, which can account for stars with higher accretion rates.
In addition, we performed new simulations with parameter variations from \mFid, as shown by the \texttt{Fid\_*} models in Fig. \ref{fig:Mdot-t_obs}.
The model \texttt{Fid\_apz=2e-3} shows that a simple doubling of $\apz$ can also lead to higher accretion rates.
When we change $\cwo$ to $5 \times 10^{-6}$ (\texttt{Fid\_Cw0=5e-6}) and $1 \times 10^{-6}$ (\texttt{Fid\_Cw0=1e-6}), then changing $\cwo$ by an order of magnitude can account for all the median points in \citet{Testi2022The-protoplanet}.
These models correspond to the case when the wind mass loss is more or less weaker than assumed in \mFid.
By considering the dispersion of the wind mass-loss rates for each object, we can therefore explain the observed $\dot{M}$--$t$ relation in wind-driven accretion disks.

The disk winds in the inner region may fail to escape the gravitational potential and directly accrete onto the star.
\citet{Takasao2022Three-dimension} showed that after disk winds are launched from the surrounding disk to the star, they fall toward the star. This may occur for winds that are launched around $r \sim 0.1$--$1 $au.

\subsection{Updates from previous migration snowlines} \label{ssec:dis-sl}

We describe updates on the snowline migration compared with those in previous studies.
In our calculations, the migration of the snowlines basically arrives at 1 au in the early stage of the Class II disk evolution.
The typical arrival time at 1 au is $t \sim$ 0.2~Myr, which is earlier than that in \citetalias{Mori2021Evolution-of-th}.
This arrival time is still the case even in \mDG, where accretion heating is enhanced by dust growth \citep{Kondo2023The-Roles-of-Du}.
One update is that the $\erad$ we chose is ten times lower than those of previous studies.
In addition, the wind mass loss may reduce the accretion rate by a factor of $\sim 2$--5.
Furthermore, in our model, we modified $t = 0 $ in the luminosity evolution to be on the birthline \citep{Stahler2004The-Formation-o}, as described in Sect. \ref{ssec:Tirr}.
This decreased the luminosity at $ t \lesssim $ 0.5 Myr and shifts the arrival time earlier.
If these early arrivals of the snowline are the case, the formation of rocky protoplanets in the Class 0--I phase may be plausible and would avoid involving icy dust in rocky protoplanets.

Moreover, by considering disk dispersal, we newly obtained the inner limit of the snowline.
The limit is about 0.3 au, independent of the accretion heating models.
It might naively be expect that if the disk is dispersed before the snowline arrives at 1 au, the Earth would not be exposed to the icy material.
However, even in the \mEff model, the snowline stays within 1 au for $\sim$ 1 Myr before the disk disperses.
If the disk dispersed suddenly, this would have to occur earlier than 1 Myr. This contradicts the observed disk lifetime.
Thus, disk dispersal would not prevent the Earth from being exposed to icy material.

\subsection{Caveats: Thermal ionization in the innermost region}
We did not take thermal ionization into account. This occurs at $T \gtrsim 1000$ K \citep[see][]{Desch2015High-temperatur}.
When thermal ionization occurs, it may increase the ionization fraction and then trigger MRI.
The MRI turbulence may cause viscous heating and then maintain the thermal ionization and turbulence \citepalias{Mori2021Evolution-of-th}.
This effect may influence the temperature structure within $\sim$ 1 au in the early phase and within $\sim$ 0.1 au in the late phase, as well as the evolution of protoplanets (Sect. \ref{ssec:infl-plf}).
The sustainability of this self-sustained MRI remains uncertain and needs to be investigated in future studies.

\section{Summary} \label{sec:summary}
We investigated the evolution of the thermal structure of magnetized disks with wind mass loss.
Our model was designed to be magnetized wind-driven accretion disks, where accretion is driven by wind stress.
The temperature model incorporated recent findings: surface heating, and energy loss by disk wind.
In addition, whereas previous studies (\citetalias{Mori2021Evolution-of-th}; \citealt{Kondo2023The-Roles-of-Du}) assumed uniform accretion rates, the accretion rate distribution is affected by the wind mass loss.
This study highlighted the importance of choosing a temperature model that is consistent with the disk dynamics and demonstrated that the difference in the temperature model impacts the formation of super-Earths and sub-Neptunes. Our findings are listed below.
\begin{itemize}

\item
Lower temperatures in MHD disks.
The temperature profile in the MHD heating model is invariably lower than in the classical heating model. The differences reach up to a factor of five.
The main contributor to the attenuation of accretion heating is the reduced effective optical depth that is due to surface heating in the MHD heating.
In addition, energy loss due to disk winds further reduces the accretion heating.
Furthermore, even in inner regions, the accretion rate is suppressed by wind mass loss, which reduces the accretion heating.

\item
Limited accretion-heated region.
The accretion-heated region is limited to within 0.1--1 au even in the early phase of Class II.
This means that the overall temperature profile is primarily shaped by irradiation heating, which allowed us to simplify models of the temperature structure and evolution of protoplanetary disks.

\item
Impact on planetary evolution and water fraction.
The differences in the temperature models significantly impact the planetary evolution and water fraction. In the MHD heating model, the pebble isolation mass is lower than in the classical heating model because the temperature is lower. In addition, as the snowline lies at inner radii in the MHD heating model, the obtained close-in planets have a dichotomy of lower-mass rocky planets and higher-mass volatile-rich planets. This dichotomy transitions around $\Mp \sim 4\,M_\oplus$. This dichotomy appears to be consistent with exoplanet observations \citep[e.g.,][]{Parc2024From-super-Eart}.

\end{itemize}

Although the stellar accretion rate in our model is suppressed by wind mass loss, this may be consistent with the observational relation of the stellar accretion rate and age (see Sect. \ref{ssec:disc-compobs}). The observational relation still has large uncertainties.

This study encourages further investigation into the earlier phases of disk evolution (i.e., Class 0--I). Whereas the disk temperature during the Class II phase may be cold, as suggested by MHD disk models, earlier disks are likely more massive due to gas infall from the cloud core \citep[e.g.,][]{Kimura2016From-Birth-to-D,Marschall2023An-inflationary}. This could trigger gravitational instability, which would lead to shock and compressional heating from gravitational spirals that efficiently heats the disk. Future studies of the evolution of the temperature structures from disk formation to disk dispersal are required.

\begin{acknowledgements}
We thank Takeru Suzuki and Xuening Bai for valuable discussions, particularly regarding the energetics of our model, and Satoshi Okuzumi for his insightful comments on the comparison with observational data.
We also thank Chris Ormel, Bertram Bitsch, Alessandro Morbidelli, Tristan Guillot, and Aur\'elien Crida, and the anonymous referee for insightful discussions and constructive feedback.
    This research has made use of the NASA Exoplanet Archive, which is operated by the California Institute of Technology, under contract with the National Aeronautics and Space Administration under the Exoplanet Exploration Program.
    S.M. is supported by the JSPS KAKENHI (grant Nos. JP21J00086, 22KJ0155, and 22K14081) and Shuimu Tsinghua Scholar program.
    M.K. is supported by the JSPS KAKENHI (grant Nos. 23H01227, 24K00654, and 24K07099) and thanks Observatoire de la C\^{o}te d'Azur for the hospitality during his long-term stay in Nice.
    M.O. is supported by the National Natural Science Foundation of China (Nos. 12273023, 12250610186).
    Numerical computations were partially carried out on a PC cluster at the Center for Computational Astrophysics, National Astronomical Observatory of Japan.
\end{acknowledgements}

\bibliographystyle{aa}

\bibliography{mybib}

\begin{appendix}

\onecolumn
\section{Pebble-to-gas flux ratio from dust transport simulation}
\label{app:pebble-twopop}

\begin{figure*}[h!]
    \centering
    \includegraphics[width=0.99\linewidth]{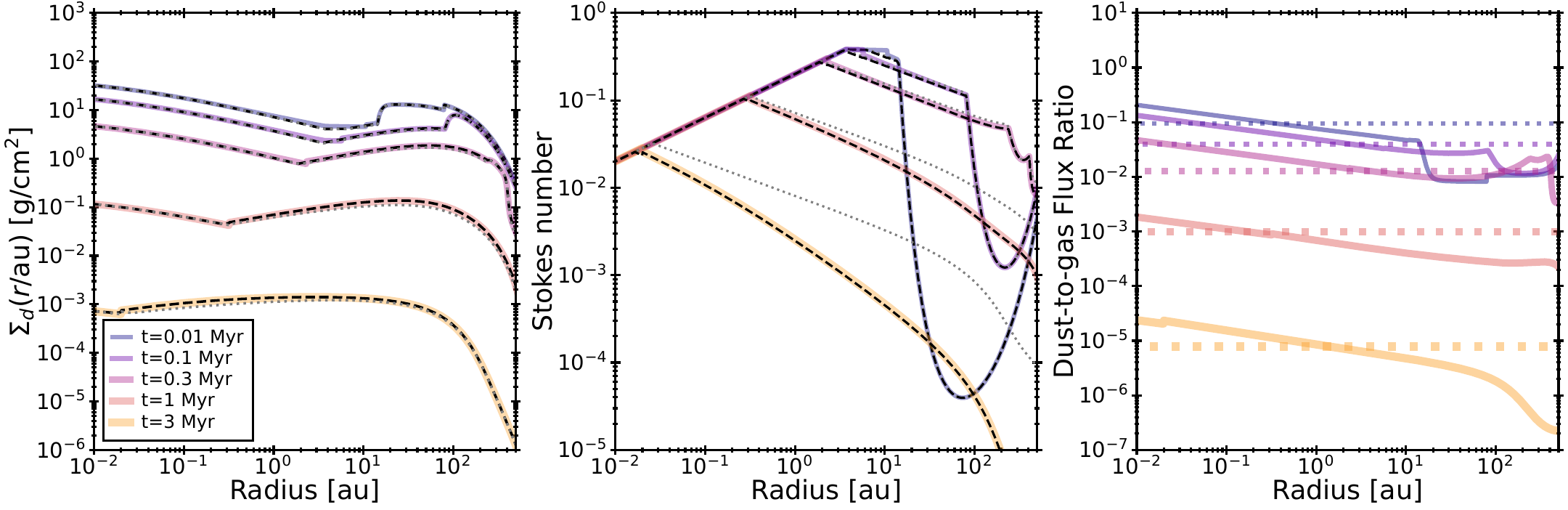}
    \caption{
    Pebble profiles obtained from a two population dust model \citep[][see Appendix \ref{app:pebble-twopop} for assumed disk profiles]{Birnstiel2012A-simple-model-}, in comparison with a prediction of the dust surface density $\Sigma_{\rm d}$ and Stokes number for largest dust grains (black dashed; Eq. (\ref{eq:St})). The classical estimate for $\St$ using Eq. (\ref{appeq:St-drift}) is also shown in the middle panel (gray dotted).
    \label{fig:pebbles}
    }
\end{figure*}

To obtain a typical value of the dust-to-gas flux ratio $\xi$ and its evolution, we calculated the dust transport based on a two-population dust model \citep{Birnstiel2012A-simple-model-}, as shown in Fig. \ref{fig:pebbles}.
We here assumed a simple gas structure: surface density profile $\Sigma = 1000\, \gcmcm (r/\au)^{-0.75}  \exp\left[ - (r/100\, \au) -(t/1\,{\rm Myr})^{0.5} \right ]$, temperature profile $T = 170\, \K \, (r/\au)^{-0.5}$, viscosity strength $\alpha_\nu = 10^{-2}$, and turbulent strength and dust diffusivity $\alpha_{\rm t} = 10^{-4}$. Other parameters are the same as in Sect. \ref{ssec:model-growth}.
Figure\,\ref{fig:pebbles} shows the simulated dust surface density $\Sigma_{\rm d}$, Stokes number $\St$, and dust-to-gas flux ratio $\xi = \dot{M}_{\rm peb}/\dot{M}_{\rm acc}$.
Although $\xi$ has the radial dependence, we here simply fitted $\xi$ at $r = 1 $ au, which gives Eq. (\ref{eq:fit-xi-eq}).
We plot the fitted $\xi$ values on the right panel of Fig. \ref{fig:pebbles}.

\section{Simple prediction of the pebble Stokes number}
\label{app:pebbleSt}

We updated the classical estimate for the Stokes number of pebbles (i.e., the largest grains) in the radial-drift limit \citep[e.g.,][]{Birnstiel2012A-simple-model-,Lambrechts2014Forming-the-cor}, which is given by
\begin{equation}
\label{appeq:St-drift}
        \mathrm{St}_{\mathrm{drift}} = \sqrt{0.55 \frac{|\dot{M}_{\mathrm{peb}}|} {8 \pi r \Sigma_g \eta^2 v_{\rm K}}}  .
\end{equation}
However, this expression may break down in the presence of strong gas advection, which determines the dust mass flux (i.e., when $|v_{\rm gas}| \gtrsim |2 \eta v_{\rm K}|$).

The upper limit of $\St$ for the drift limit is obtained by equating the growth timescale with the drift timescale of pebbles.
The timescale of pebble drift on a gas comoving frame is written as $ \sim r / | - 2 \eta v_{\rm K} \St |$.
Meanwhile, the growth timescale is $\sim (\Omega \Sigma_{\rm d} / \Sigma_{\rm g} )^{-1} $.
Thus, we obtained
\begin{equation}
\label{app-eq:timescale}
        r / | - 2 \eta v_{\rm K} \St | \sim  (\Omega \Sigma_{\rm peb} / \Sigma_{\rm g} )^{-1}.
\end{equation}
The pebble mass flux $\dot{M}_{\rm peb}$ is written as
\begin{equation}
\label{app-eq:Mdotpeb}
        \dot{M}_{\rm peb} = - 2 \pi r \Sigma_{\rm peb} (- 2 \eta v_{\rm K} \St + v_{\rm gas})
\end{equation}
 for $\St < 1$, although some studies here assumed $|v_{\rm gas}| \ll |2 \eta v_{\rm K}|$.
Combining Eqs. (\ref{app-eq:timescale}) and (\ref{app-eq:Mdotpeb}), we have a quadratic equation of $ \St$ and obtain
\begin{equation}
        \mathrm{St}_{\mathrm{drift}} =  \tilde{v}_{\mathrm{gas} }
        + \sqrt{\tilde{v}_{\mathrm{gas}}^2 + 0.55\, \frac{\dot{M}_{\mathrm{peb}}} {8\pi  \Sigma_g r v_{\rm K} \eta | \eta |}} ,
\end{equation}
where $\tilde{v}_{\mathrm{gas}} \equiv v_{\mathrm{gas}} /(4 \eta v_{\rm K})$ and we considered the fitting parameter of \citet{Birnstiel2012A-simple-model-}.
The key point is that the drift timescale should be measured in the gas comoving frame, whereas the pebble mass flux must include the background flow.
Once we obtain $\St$, we can calculate the pebble surface density from Eq. (\ref{app-eq:Mdotpeb}).

To demonstrate the obtained formulae, we overplot the analytical expressions for the Stokes number (Eq. (\ref{eq:St})) and the resulting pebble surface density on the left and middle panels of Fig. \ref{fig:pebbles}, respectively.
Here, the pebble mass flux was taken from the simulation.
We confirmed that the formulae accurately reproduce the results of the dust transport simulation.
In contrast, the prediction of $\St$ using Eq. (\ref{appeq:St-drift}) deviates from the simulation in the outer region at $t \gtrsim 1$ Myr.
Note that we adopted a high viscosity in this calculation, which leads to a high gas velocity, causing it to exceed the drift velocity expected from the gas motion.

\twocolumn

\onecolumn
\section{Dependence of planetary evolution path on parameters}
\label{app:tracks}

We present planetary evolution tracks for parameter sets other than those in Sect. \ref{ssec:infl-plf}:
a case with a lower fragmentation velocity ($v_{\rm frag} = 1$ m s$^{-1}$) and a case with later onset of protoplanet growth ($t_0 = 0.1$  Myr).

\begin{figure*}[h!]
\fboxsep = 1.pt
        \centering \figtitlen{0.2}{0.12}
    \includegraphics[height=0.35\linewidth, trim={0 0mm 0 0mm},clip]{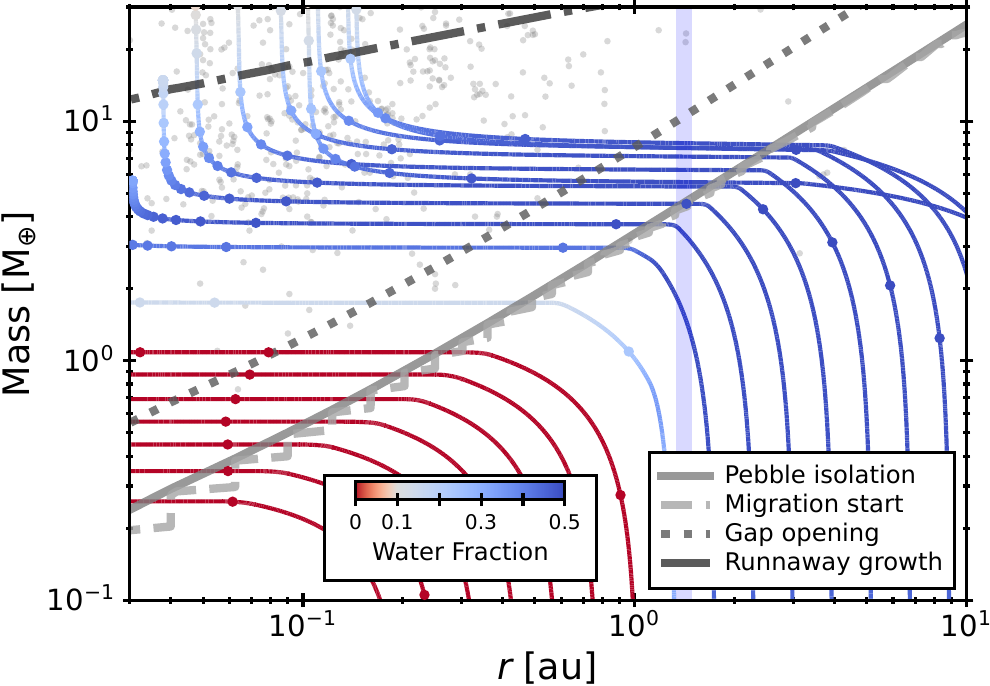}
    \includegraphics[height=0.35\linewidth, trim={20mm 0mm 0 0mm},clip]{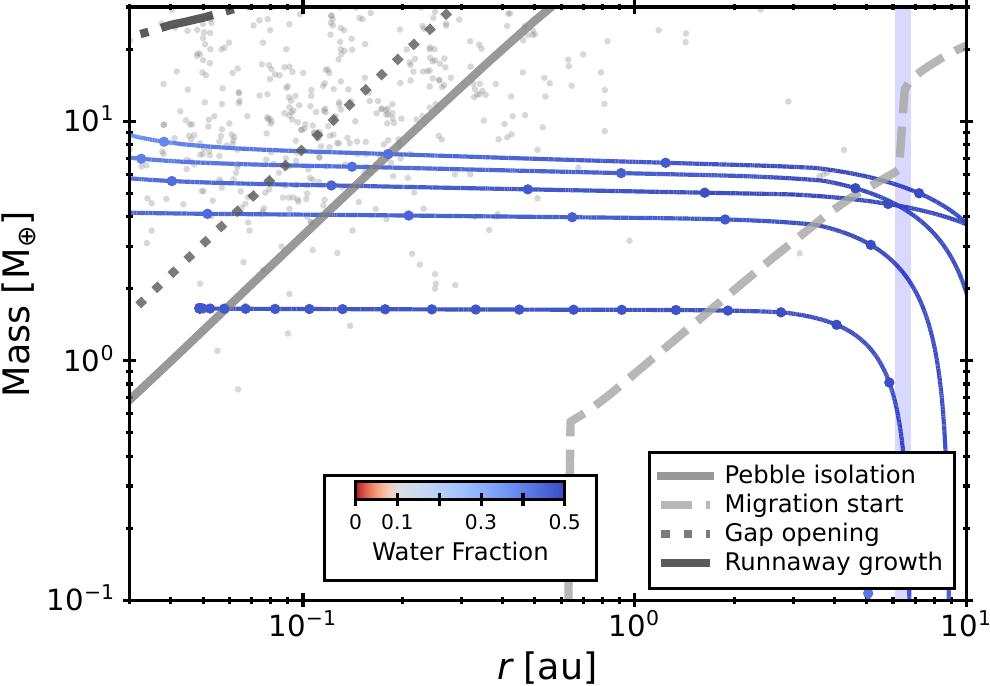}
        \vspace*{-41ex}
\begin{center}
      \hspace{0.41\linewidth} \textsf{\large \bfseries \colorbox{white}{(a)}}
      \hspace{0.41\linewidth}  \textsf{\large \bfseries \colorbox{white}{(b)}}
      \end{center}
\vspace*{29.5ex}
        \includegraphics[height=0.35\linewidth, trim={0 0 0 0mm},clip]{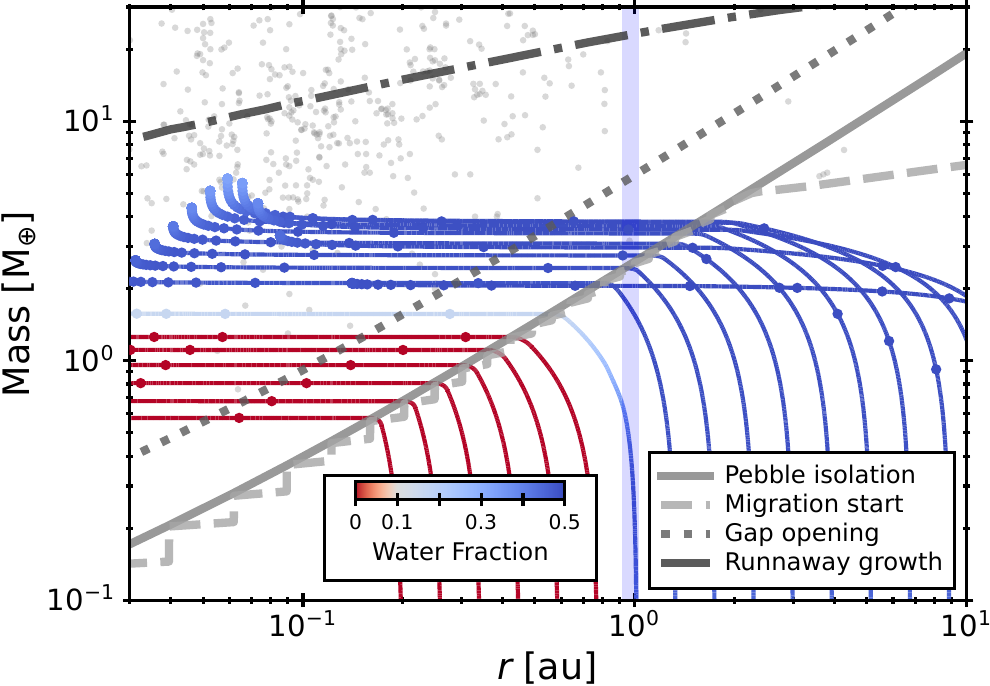}
    \includegraphics[height=0.35\linewidth, trim={20mm 0 0 0mm},clip]{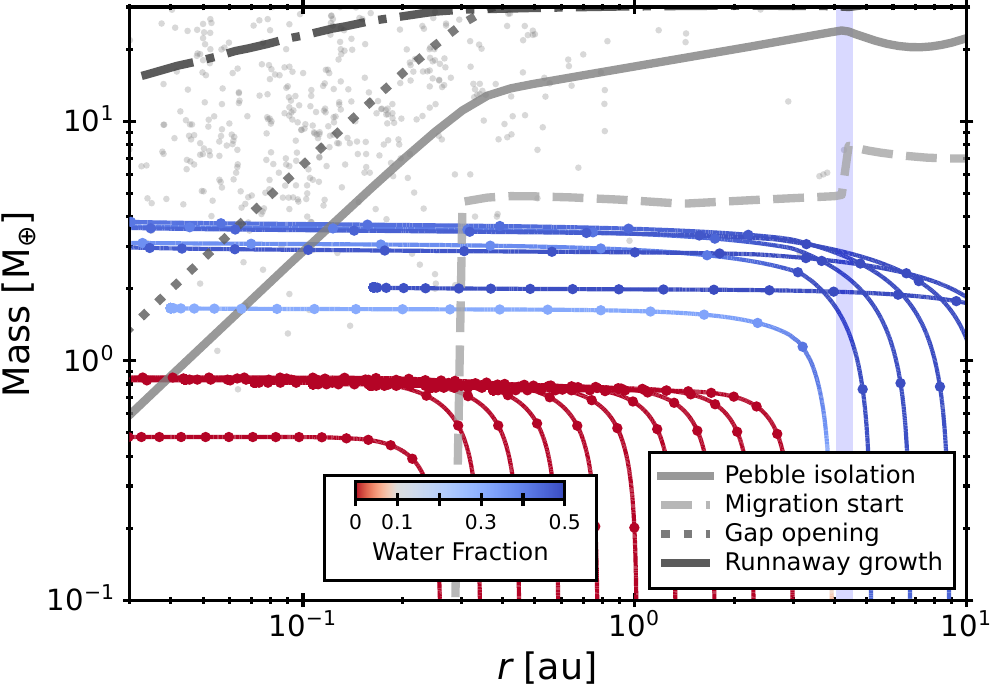}
    \vspace*{-41ex}
\begin{center}
      \hspace{0.41\linewidth} \textsf{\large \bfseries  \colorbox{white}{(c)}}
      \hspace{0.41\linewidth}  \textsf{\large \bfseries  \colorbox{white}{(d)}}
      \end{center}
\vspace*{35ex}
   \caption{Same as Fig. \ref{fig:masses} but for different parameter sets: (top) a case with $t_0 = 0.01$ Myr and $v_{\rm frag} = 1$ m s$^{-1}$, and (bottom) a case with $t_0 = 0.2$ Myr and $v_{\rm frag} = 10$ m s$^{-1}$. The total drifting pebble masses at the snowline for the former and later cases are 400 $M_\oplus$ and 100 $M_\oplus$, respectively.
   }
    \label{fig:app-tracks}
\end{figure*}

\end{appendix}

\end{document}